\documentclass{article}
\usepackage{amsmath}
\usepackage{graphicx}
\usepackage{amsfonts}
\usepackage{amsthm}
\usepackage{times}
\usepackage{fancyhdr}
\providecommand{\ve}[1]{\boldsymbol{#1}}

\providecommand{\norm}[1]{\left \lVert#1 \right  \rVert}

\newcommand{\T}{^{\ensuremath{\mathsf{T}}}}           
\DeclareMathOperator*{\argmax}{argmax}
\DeclareMathOperator*{\argmin}{argmin}

\newtheorem{thm}{Theorem}
\newcommand{\thma}{\begin{thm}}
\newcommand{\thmb}{\end{thm}}

\newtheorem{defi}{Definition}
\newcommand{\defa}{\begin{defi}}
\newcommand{\defb}{\end{defi}}

\newcommand{\en}{\end{equation}}
\newcommand{\eqb}{\end{equation}}

\newcommand{\enuma}{\begin{enumerate}}
\newcommand{\enumb}{\end{enumerate}}
\newcommand{\ena}{\begin{enumerate}}
\newcommand{\enb}{\end{enumerate}}

\newcommand{\itema}{\begin{itemize}}
\newcommand{\itemb}{\end{itemize}}
\newcommand{\ita}{\begin{itemize}}
\newcommand{\itb}{\end{itemize}}

\newcommand{\alignb}{\end{align}}

\newcommand{\proofa}{\begin{proof}}
\newcommand{\proofb}{\end{proof}}

\newcommand{\bla}{\begin{block}}
\newcommand{\blb}{\end{block}}

\newcommand{\seqb}{\end{equation*}}

\newcommand{\bth}{\ve{\theta}}
\newcommand{\thetn}{\ve{\theta}}
\newcommand{\thet}{\thetn}

\newcommand{\vth}{\ve{\thet}}

\newcommand{\hbth}{\widehat{\thet}}

\newcommand{\Ca}{[\text{Ca}^{2+}]}

\newcommand{\bd}{\ve{d}}

\newcommand{\bg}{\ve{g}}

\newcommand{\bn}{\ve{n}}

\newcommand{\bC}{\ve{C}}

\newcommand{\bF}{\ve{F}}

\newcommand{\bH}{\ve{H}}
\newcommand{\bI}{\ve{I}}

\newcommand{\bM}{\ve{M}}

\newcommand{\bX}{\ve{X}}

\newcommand{\vF}{\vec{F}}

\newcommand{\hn}{\widehat{n}}

\newcommand{\hC}{\widehat{C}}

\newcommand{\hbn}{\widehat{\ve{n}}}

\newcommand{\hbC}{\widehat{\ve{C}}}

\newcommand{\vbF}{\vec{\ve{F}}}

\newcommand{\mN}{\mathcal{N}}

\newcommand{\mP}{\mathcal{P}}

\newcommand{\Del}{\Delta}
\newcommand{\Sig}{\Sigma}

\newcommand{\sig}{\sigma}
\newcommand{\lam}{\lambda}
\newcommand{\gam}{\gamma}

\newcommand{\balpha}{\ve{\alpha}}
\newcommand{\bbeta}{\ve{\beta}}

\newcommand{\blam}{\ve{\lambda}}

\newcommand{\bGam}{\ve{\Gamma}}

\newcommand{\valpha}{\vec{\alpha}}
\newcommand{\vbeta}{\vec{\beta}}

\newcommand{\vbalpha}{\vec{\balpha}}

\newcommand{\hbalpha}{\widehat{\ve{\alpha}}}
\newcommand{\hbbeta}{\widehat{\ve{\hbeta}}}

\newcommand{\halpha}{\widehat{\alpha}}
\newcommand{\hbeta}{\widehat{\beta}}

\newcommand{\hsig}{\widehat{\sigma}}
\newcommand{\hlam}{\widehat{\lambda}}

\usepackage{multicol}
\usepackage{hyperref}
\usepackage{algorithmic}
\usepackage{algorithm}
\newcommand{\zzz}{z}
\newcommand{\az}{\argmax_{\bM \bC \geq \ve{0}}}
\newcommand{\anx}{\argmax_{n_t \in \mathbb{N}_0 \forall t}}
\newcommand{\foopsi}{fast }

\title{Fast non-negative deconvolution for spike train inference from population calcium imaging}

\author{Joshua T.~Vogelstein, Adam M.~Packer, Tim A.~Machado, \\ Tanya Sippy, Baktash Babadi, Rafael Yuste, Liam Paninski}

\begin{document}

\maketitle
\begin{abstract}
Calcium imaging for observing spiking activity from large populations of neurons are quickly gaining popularity.  While the raw data are fluorescence movies, the underlying spike trains are of interest.  This work presents a fast non-negative deconvolution filter to infer the approximately most likely spike train for each neuron, given the fluorescence observations. This algorithm outperforms optimal linear deconvolution (Wiener filtering) on both simulated and biological data. The performance gains come from restricting the inferred spike trains to be positive (using an interior-point method), unlike the Wiener filter.  The algorithm is fast enough that even when imaging over 100 neurons, inference can be performed on the set of all observed traces faster than real-time.  Performing optimal spatial filtering on the images further refines the estimates.  Importantly, all the parameters required to perform the inference can be estimated using only the fluorescence data, obviating the need to perform joint electrophysiological and imaging calibration experiments.
\end{abstract}

\section{Introduction}


Simultaneously imaging large populations of neurons using calcium sensors is becoming increasingly popular \cite{ImagingManual}, both \emph{in vitro} \cite{SmettersYuste99, IkegayaYuste04} and \emph{in vivo} \cite{NagayamaChen07, GobelHelmchen07, LuoSvoboda08}, and will likely continue as the signal-to-noise-ratio (SNR) of genetic sensors continues to improve \cite{GaraschukKonnerth07, MankGriesbeck08b, WallaceHasan08}. 
Whereas the data from these experiments are movies of time-varying fluorescence traces, the desired signal consists of spike trains of the observable neurons. Unfortunately, finding the most likely spike train is a challenging computational task, due to limitations of the SNR and temporal resolution, unknown parameters, and computational intractability. 

A number of groups have therefore proposed algorithms to infer spike trains from calcium fluorescence data using very different approaches.  Early approaches simply thresholded $dF/F$ (e.g., \cite{Schwartz98,Mao01}) to obtain ``event onset times.''  More recently, Greenberg et al. \cite{GreenbergKerr08} developed a  template matching algorithm to identify individual spikes. 
Holekamp et al. \cite{HolekampHoly08} then applied an optimal linear deconvolution (ie, the Wiener filter) to the fluorescence data.  This approach is natural from a signal processing standpoint, but does not utilize the knowledge that spikes are always positive.  Sasaki et al. \cite{SasakiIkegaya08} proposed using machine learning techniques to build a nonlinear supervised classifier, requiring many hundreds of examples of joint electrophysiological and imaging data to ``train'' the algorithm to learn what effect spikes have on fluorescence.  Vogelstein et al. \cite{VogelsteinPaninski09} proposed a biophysical model-based sequential Monte Carlo method to efficiently estimate the probability of a spike in each image frame, given the entire fluorescence time-series.  While effective, that approach is not suitable for online analyses of populations of neurons, as the computations run in about real-time per neuron (ie, analyzing one minute of data requires about one minute of computational time on a standard laptop computer).

The present work starts by building a simple model relating spiking activity to fluorescence traces. Unfortunately, inferring the most likely spike train given this model is computationally intractable.  Making some reasonable approximations leads to an algorithm that infers the approximately most likely spike train, given the fluorescence data.  This algorithm has a few particularly noteworthy features, relative to other approaches.  First, spikes are assumed to be positive.  This assumption often improves filtering results when the underlying signal has this property \cite{PortugalVicente94, MarkhamConchello99, LeeSeung99, LLS04, OGradyPearlmutter06, HuysPaninski06, Cunningham08, PaninskiWu09}.  Second, the algorithm is fast: it can process a calcium trace from 50,000 images in about one second on a standard laptop computer. In fact, filtering the signals for an entire population of about $100$ neurons runs faster than real-time. This speed facilitates using this filter online, as observations are being collected. In addition to these two features, the model may be generalized in a number of ways, including incorporating spatial filtering of the raw movie. The efficacy of the proposed filter is demonstrated on several biological data-sets, suggesting that this algorithm is a powerful and robust tool for online spike train inference.  The code (which is a simple Matlab script) is available from the authors upon request.

\section{Methods} \label{sec:methods}


\subsection{Data driven generative model} \label{sec:model}

\begin{figure}[h!]
\centering \includegraphics[width=.9\linewidth]{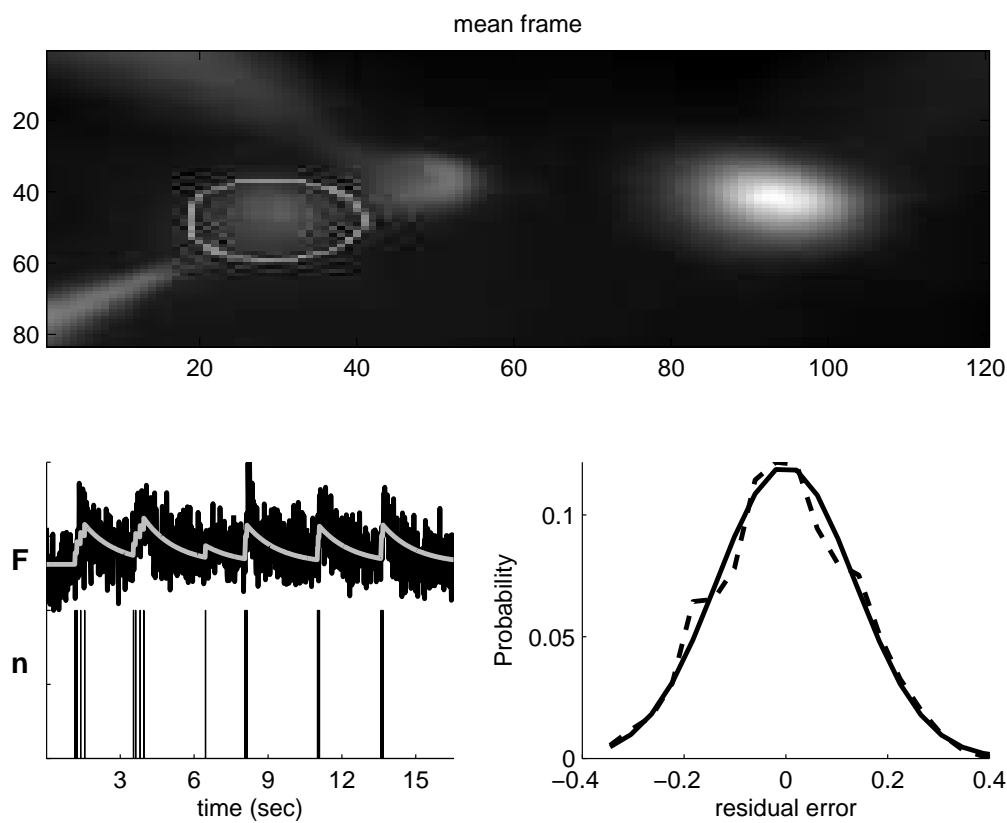}
\caption[data-based model]{Typical \emph{in vitro} data suggest that a reasonable first order model may be constructed by convolving the spike train with an exponential and adding Gaussian noise. Top panel: the average (over frames) of a typical field-of-view.  Bottom left: true spike train recorded via a patch electrode (black bars), convolved with an exponential (gray line), superimposed on the fluorescence trace (OGB-1; black line).  While the spike train and fluorescence trace are measured data, the calcium is not directly measured, but rather, inferred.  Bottom right: a histogram of the residual error between the gray and black lines from the bottom left panel (dashed line), and the best fit Gaussian (solid line). Note that the Gaussian model provides a good fit for the residuals here.} \label{fig:in_vitro_ex}
\end{figure}

Figure \ref{fig:in_vitro_ex} shows data from a typical \emph{in vitro} epifluorescence experiment (see section \ref{sec:exp} for data collection details).  The top panel shows the mean frame of this movie, including 3 neurons, two of which are patched.  To build the model, the pixels within a region-of-interest (ROI) are selected (white circle).  Given the ROI, all the pixel intensities of each frame can be averaged, to get a one-dimensional fluorescence time-series, as shown in the bottom left panel (black line).  By patching onto this neuron, the spike train can also be directly observed (black bars). Previous work suggests that this fluorescence signal might be well characterized by convolving the spike train with an exponential, and adding noise \cite{ImagingManual}.  This model is confirmed by convolving the true spike train with an exponential (gray line, bottom left panel), and then looking at the distribution of the residuals.  The bottom right panel shows a histogram of the residuals (solid line), and the best fit Gaussian distribution (dashed line).

The above observations may be formalized as follows. Assume there is a one-dimensional fluorescence trace, $\bF$ (throughout this text $\bX$ indicates the vector $[X_1, \ldots, X_T]$, where $T$ is the index of the final frame), from a neuron.  At time $t$, the fluorescence measurement $F_t$ is a linear-Gaussian function of the intracellular calcium concentration at that time, $\Ca_t$:
\begin{align} \label{eq:F}
F_t &= \alpha (\Ca_t + \beta) + \sig \varepsilon_t, \qquad \varepsilon_t \overset{iid}{\sim} \mN(0,1).
\end{align}
\noindent The scale, $\alpha$, absorbs all experimental variables impacting the scale of the signal, including the number of sensors within the cell, photons per calcium ion, amplification of the imaging system, etc.  Similarly, the offset, $\beta$, absorbs the baseline calcium concentration of the cell, background fluorescence of the fluorophore, imaging system offset, etc.  The standard deviation, $\sig$, results from calcium fluctuations independent of spiking activity, fluorescence fluctuations independent of calcium, and imaging noise. The noise at each time, $\varepsilon_t$, is independently and identically distributed according to a standard normal distribution (ie, Gaussian with zero mean and unit variance), as indicated by the notation $\overset{iid}{\sim}\mN(0,1)$. 

Then, assuming that the intracellular calcium concentration, $\Ca_t$, jumps by $A$ $\mu$M after each spike, and subsequently decays back down to $C_b$ $\mu$M with time constant $\tau$, one can write:
\begin{align} \label{eq:C1}
\Ca_{t+1} = (1- \Del/\tau ) \Ca_t + (\Del/\tau) C_b + A n_t
\end{align}
\noindent where $\Del$ is the time step size --- which is the frame duration, or $1/$(frame rate) --- and $n_t$ indicates the number of times the neuron spiked in frame $t$. 
Note that because $\Ca_t$ and $F_t$ are linearly related to one another, the fluorescence scale, $\alpha$, and calcium scale, $A$, are not identifiable.  In other words, either can be set to unity without loss of generality, as the other can absorb the scale entirely. Similarly, the fluorescence offset, $\beta$, and calcium baseline, $C_b$ are not identifiable, so either can be set to zero without loss of generality.  Finally, letting $\gam=(1-\Del/\tau)$, Eq.~\eqref{eq:C1} can be rewritten replacing $\Ca_t$ with its non-dimensionalized counterpart, $C_t$: 
\begin{align} \label{eq:C2}
	 C_{t+1}=\gam C_t + n_t.
\end{align} 
\noindent Note that $C_t$ does not refer to absolute intracellular concentration of calcium, but rather, a relative measure (see \cite{VogelsteinPaninski09} for a more general model).  The gray line in the bottom left panel of Figure \ref{fig:in_vitro_ex} corresponds to the putative $\bC$ of the observed neuron.  

To complete the ``generative model'' (ie, a model from which simulations can be generated), the distribution from which spikes are sampled must be defined.  Perhaps the simplest first order description of spike trains is that at each time, spikes are sampled according to a Poisson distribution with some rate:
\begin{align} \label{eq:n}
	n_t \overset{iid}{\sim} \text{Poisson}(\lam \Del)
\end{align}
\noindent where $\lam \Del$ is the expected firing rate per bin, and $\Del$ is included to ensure that the expected firing rate is independent of the frame rate.  Thus, Eqs.~\eqref{eq:F}, \eqref{eq:C2}, and \eqref{eq:n} complete the generative model.

\subsection{Goal} \label{sec:goal}

Given the above model, the goal is to find the maximum \emph{a posteriori} (MAP) spike train, i.e.,  the most likely spike train, $\hbn$,  given the fluorescence measurements, $\bF$:
\begin{align} \label{eq:nhat1} 
\hbn &=  \anx P[\bn | \bF], 
\end{align}
\noindent where $P[\bn | \bF]$ is the posterior probability of a spike train, $\bn$, given the fluorescent trace, $\bF$, and $n_t$ is constrained to be an integer, $\mathbb{N}_0=\{0,1,2,\ldots\}$, because of the above assumed Poisson distribution.  From Bayes' rule, the posterior can be rewritten:
\begin{align} \label{eq:bayes}
P[\bn | \bF] = \frac{P[\bn, \bF]}{P[\bF]} = \frac{1}{P[\bF]} P[\bF | \bn] P[\bn],
\end{align}
\noindent where $P[\bF]$ is the evidence of the data, $P[\bF | \bn]$ is the likelihood of observing a particular fluorescence trace $\bF$, given the spike train $\bn$, and $P[\bn]$ is the prior probability of a spike train.  Plugging the far right-hand-side of Eq.~\eqref{eq:bayes} into Eq.~\eqref{eq:nhat1}, yields:
\begin{align} \label{eq:nhat2} 
\hbn &=  \anx \frac{1}{P[\bF]} P[\bF | \bn] P[\bn] =  \anx  P[\bF | \bn] P[\bn],
\end{align}
\noindent where the second equality follows because $P[\bF]$ merely scales the results, but does not change the relative quality of various spike trains.  Both $P[\bF | \bn]$ and $P[\bn]$ are available from the above model:
\begin{subequations} \label{eq:post1}
\begin{align}
P[\bF | \bn]&= P[\bF | \bC] 	= \prod_{t=1}^T P[F_t | C_t], \label{eq:lik1} \\ 
P[\bn] 		&= \prod_{t=1}^T P[n_t], \label{eq:prior1}
\end{align}
\end{subequations}
\noindent where the first equality in Eq.~\eqref{eq:lik1} follows because $\bC$ is deterministic given $\bn$, and the second equality follows from Eq.~\eqref{eq:F}. Further, Eq.~\eqref{eq:prior1} follows from the Poisson process assumption, Eq.~\eqref{eq:n}.  Both $P[F_t | C_t]$ and $P[n_t]$ can be written explicitly:
\begin{subequations} \label{eq:post2}
\begin{align}
P[F_t | C_t] &= \mN(\alpha(C_t+\beta),\sig^2), \label{eq:lik2} \\
P[n_t] &= \text{Poisson}(\lam \Del), \label{eq:prior2} 
\end{align}
\end{subequations}
where both equations follow from the above model.  Now, plugging Eq.~\eqref{eq:post2} back into \eqref{eq:post1}, and plugging that result into Eq.~\eqref{eq:nhat2}, yields:
\begin{subequations}  \label{eq:obj}
\begin{align}
\hbn 	&= \anx \prod_{t=1}^T \frac{1}{\sqrt{2 \pi \sig^2}} \exp \left\{-\frac{1}{2}\frac{(F_t - \alpha (C_t + \beta))^2}{\sig^2}\right\} \frac{\exp\{-\lam\Del\} (\lam\Del)^{n_t}}{n_t!}
\label{eq:obj1}\\ &= \anx  \sum_{t=1}^T \left\{ -\frac{1}{2 \sig^2}(F_t - \alpha(C_t + \beta))^2  +  n_t \log \lam \Del - \log n_t! \right\}, \label{eq:logobj1}
\end{align} 
\end{subequations}
\noindent where the second equality follows from taking the logarithm of the right-hand-side and dropping terms that do not depend on $\bn$.  Unfortunately, solving Eq.~\eqref{eq:logobj1} exactly is computationally intractable, as it requires a nonlinear search over an infinite number of  possible spike trains.  The search space could be restricted by imposing an upper bound, $k$, on the number of spikes within a frame.  However, in that case, the computational complexity scales \emph{exponentially} with the number of image frames --- i.e.,  the number of computations required would scale with $k^T$ --- which for pragmatic reasons is intractable.

\subsection{Inferring the most likely spike train, given a fluorescence trace} \label{sec:inf}

The goal here is to develop an algorithm to efficiently approximate $\hbn$, the most likely spike train, given the fluorescence trace. Because of the computational intractability described above, Eq.~\eqref{eq:obj} is approximated by modifying Eq.~\eqref{eq:n}, replacing the Poisson distribution with an exponential distribution of the same mean. Modifying Eq.~\eqref{eq:obj} to incorporate this approximation yields:
\begin{subequations}
\begin{align} \label{eq:obj2}
\hbn &\approx \argmax_{n_t>0 \, \forall t} \prod_{t=1}^T  \left\{\frac{1}{\sqrt{2 \pi \sig^2}} \exp \left\{-\frac{1}{2}\frac{(F_t - \alpha (C_t + \beta))^2}{\sig^2}\right\}  (\lam\Del) \exp\{-\lam\Del n_t\}\right\}
\\ &= \argmax_{n_t>0 \, \forall t}  \sum_{t=1}^T -\frac{1}{2 \sig^2}(F_t - \alpha(C_t + \beta))^2  - n_t \lam \Del  \label{eq:obj3}
\end{align}
\end{subequations}
where the constraint on $n_t$ has been relaxed from  $n_t \in \mathbb{N}_0$ to $n_t \geq 0$ (since the exponential distribution can yield any non-negative number).  The advantage of this approximation is that the optimization problem becomes concave in $\bC$, meaning that any gradient ascent method guarantees achieving the global maximum (because there are no local maxima, other than the single global maximum).  To see that Eq.~\eqref{eq:obj3} is concave in $\bC$, rearrange Eq.~\eqref{eq:C2} to obtain, $n_t=C_t-\gam C_{t-1}$, so Eq.~\eqref{eq:obj3} can be rewritten:
\begin{align}
\bC &= \argmax_{C_t-\gam C_{t-1}>0 \, \forall t}  \sum_{t=1}^T -\frac{1}{2 \sig^2}(F_t - \alpha(C_t + \beta))^2  - (C_t -\gam C_{t-1}) \lam \Del  \label{eq:obj4}
\end{align}
\noindent which is a sum of terms that are concave in $C_t$, so the whole right-hand-side is concave. Unfortunately, the integer constraint has been lost, i.e.,  the answer could include ``partial'' spikes.  This disadvantage can be remedied by thresholding (ie, setting $n_t=1$ for all $n_t$ greater than some threshold, and the rest setting to zero), or by considering the magnitude of a partial spike at time $t$ as an indication of the probability of a spike occurring during frame $t$. Note that replacing a Poisson with an exponential is a common approximation technique in the machine learning literature \cite{CONV04, PaninskiWu09}, as the exponential distribution is the closest log-concave relaxation to its non-log-concave counterpart, the Poisson distribution. More specifically, the probability mass function of a Poisson distributed random variable with low rate is very similar to the probability density function of a random variable with an exponential distribution. While this convex relaxation makes the problem tractable, the ``sharp'' threshold imposed by the non-negativity constraint prohibits the use of standard gradient ascent techniques. This may be rectified by dropping the sharp threshold, and adding a barrier term which must approach $-\infty$ as $n_t$ approaches zero (this approach is often called an ``interior-point'' method). Iteratively reducing the weight of the barrier term guarantees convergence to the correct solution.  Thus, the goal is to efficiently solve:
\begin{align} \label{eq:eta}
\hbC_{\zzz} &= \argmax_{\bC}  \sum_{t=1}^T \left( -\frac{1}{2 \sig^2}(F_t - \alpha(C_t + \beta))^2  -  (C_t-\gam C_{t-1})  \lam \Del + \zzz \log (C_t-\gam C_{t-1}) \right).
\end{align}
\noindent where $\log (\cdot)$ is the ``barrier term'', and $z$ is the weight of the barrier term.  Iteratively solving for $\hbC_{\zzz}$ for $z$ going from one down to nearly zero, guarantees convergence to $\hbC$ \cite{CONV04}. 
The concavity of Eq.~\eqref{eq:eta} facilitates utilizing any number of techniques guaranteed to find the global maximum.  Because the argument of Eq.~\eqref{eq:eta} is twice analytically differentiable, one can use the Newton-Raphson technique \cite{Press92}. The special tridiagonal structure of the Hessian enables each Newton-Raphson step to be very efficient (as described below).  To proceed, Eq.~\eqref{eq:eta} can be rewritten in matrix notation.  Note that:
\begin{align} \label{eq:M}
\ve{M} \bC = 
\begin{bmatrix}
-\gam & 1 & 0 & 0 & \cdots & 0 \\
0 & -\gam & 1 & 0 & \cdots  & 0 \\
\vdots & \ddots & \ddots & \ddots & \ddots & \vdots  \\
0 & \cdots & 0  & -\gam & 1 & 0 \\
0 & \cdots & 0 & 0 & -\gam & 1
\end{bmatrix}
\begin{bmatrix}
C_1 \\ C_2 \\  \vdots \\ C_{T-1} \\ C_T
\end{bmatrix}
= 
\begin{bmatrix}
n_1 \\ n_2 \\ \vdots  \\ n_{T-1}
\end{bmatrix}
, 
\end{align}
\noindent where $\ve{M} \in \mathbb{R}^{(T-1) \times T}$ is a bidiagonal matrix.  Then, letting $\ve{1}$ be a $T-1$ dimensional column vector and $\blam=\lam \Del \ve{1}$ yields the objection function, Eq.~\eqref{eq:eta}, in more compact matrix notation:
\begin{align} 
\hbC_{\zzz} 
&= \az  -\frac{1}{2 \sig^2} \norm{\bF - \alpha (\bC +\beta)}_2^2 - (\bM \bC )\T \blam  + \zzz \log(\bM \bC)\T\ve{1},  \label{eq:eta3}
\end{align}
\noindent where $\bM \bC \geq \ve{0}$ indicates that every element of $\bM \bC$ is greater than or equal to zero, $\T$ indicates the transpose operation, $\log(\cdot)$ indicates an element-wise logarithm, and $\norm{x}_2$ is the standard $L_2$ norm, i.e., $\norm{x}_2^2=\sum_i x_i^2$. When using Newton-Raphson to ascend a surface, one iteratively computes both the gradient (first derivative) and Hessian (second derivative) of the argument to be maximized, with respect to the variables of interest ($\bC$ here).  Then, the estimate is updated using $\bC_z \leftarrow \bC_z + s \bd$, where $s$ is the step size and $\bd$ is the step direction obtained by solving $\bH \bd = \bg$.  The gradient, $\bg$, and Hessian, $\bH$, for this model, with respect to $\bC$, are given by:
\begin{subequations} \label{eq:NR}
\begin{align}
\ve{g} &= -\frac{\alpha}{\sig^2}(\bF -\alpha({\bC\T} + \beta)) + \ve{M}\T\blam - \zzz \ve{M}\T (\ve{M} \bC)^{-1} \label{eq:g} \\
\ve{H} &= \frac{\alpha^2}{\sig^2} \ve{I} + \zzz \ve{M}\T (\ve{M} \bC)^{-2} \ve{M} \label{eq:H}
\end{align}
\end{subequations}
\noindent where the exponents on the vector $\bM \bC$ indicate element-wise operations. The step size, $s$, is found using ``backtracking linesearches'', which finds the maximal $s$ that increases the posterior and is between zero and one \cite{Press92}.

Typically, implementing Newton-Raphson requires inverting the Hessian, i.e.,  solving $\bd = \bH^{-1} \bg$, a computation that scales \emph{cubically} with $T$ (requires on the order of $T^3$ operations). Already, this would be a drastic improvement over the most efficient algorithm assuming Poisson spikes, which would require $k^T$ operations (where $k$ is the maximum number of spikes per frame).  Here, because $\ve{M}$ is bidiagonal, the Hessian is tridiagonal, so the solution may be found in about $T$ operations, via standard banded Gaussian elimination techniques (which can be implemented efficiently in Matlab using $\bH \backslash \bg$, assuming $\bH$ is represented as a sparse matrix) \cite{PaninskiWu09}. In other words, the above approximation and inference algorithm reduces computations from \emph{exponential} time to \emph{linear} time.  Appendix \ref{sec:pseudo} contains pseudocode for this algorithm, including learning the parameters, as described below.

\subsection{Learning the parameters} \label{sec:learn}

We assumed above that the parameters governing the model, $\vth=\{\alpha, \beta, \sig, \gam, \lam\}$, were known, but in practice they are typically unknown. An algorithm to estimate the most likely parameters, $\hbth$, could proceed as follows: (i) initialize some estimate of the parameters, $\hbth$, then (ii) recursively compute $\hbn$ using those parameters, and update $\hbth$ given the new $\hbn$, until some convergence criteria is met.  Below, details are provided for each step.

\subsubsection{Initializing the parameters} \label{sec:init}

Because the model introduced above is linear, the scale of $\bF$ relative to $\bn$ is arbitrary.  Therefore, before filtering, $\bF$ is linearly ``squashed'' between zero and one, ie $\bF \leftarrow (\bF - F_{min})/(F_{max}-F_{min})$, where $F_{min}$ and $F_{max}$ are the observed minimum and maximum of $\bF$, respectively.  Given this normalization, $\alpha$ is set to one.  Because spiking is assumed to be sparse, $\bF$ tends to be around baseline, so $\beta$ is initialized to be the median of $\bF$, and $\sig$ is initialized as the median absolute deviation of $\bF$, i.e.,  $\sig=$ median$_t$($|F_t-$median$_s(F_s)|$)$/K$, where median$_i(X_i)$ indicates the median of $X$ with respect to index $i$, and $K=1.4785$ is the correction factor when using median absolute deviation as a robust estimator of the standard deviation.  Because in these data, the posterior tends to be relatively flat along the $\gam$ dimension, i.e.,  large changes in $\gam$ result in relatively small changes in the posterior, estimating $\gam$ is difficult.  Further, previous work has shown that results are somewhat robust to minor variations in time constant \cite{YaksiFriedrich06}; therefore $\gam$ is initialized at $1-\Del/(1 \text{sec})$, which is fairly typical \cite{PologrutoSvoboda04}. Finally, $\lam$ is initialized at $1$ Hz, which is between typical baseline and evoked spike rate for these data.

\subsubsection{Estimating the parameters given $\widehat{\mathbf{n}}$} \label{sec:242}

Ideally, one could integrate out the hidden variables, to find the most likely parameters:
\begin{align} \label{eq:par1}
\hbth &= \argmax_{\bth} \int P[\bF, \bC | \bth] d\bC  = \argmax_{\bth} \int P[\bF | \bC; \bth] P[\bC | \bth] d\bC.
\end{align}
However, evaluating those integrals is not particularly tractable.
Therefore, Eq.~\eqref{eq:par1} is approximated by simply maximizing the parameters given the MAP estimate of the hidden variables:
\begin{align} \label{eq:par2}
\hbth &\approx \argmax_{\bth} P[\bF, \hbC | \bth] = \argmax_{\bth} P[\bF| \hbC; \bth] P[\hbn | \bth] = \argmax_{\bth} \{ \log P[\bF | \hbC; \bth] + \log P[\hbn | \bth] \}, 
\end{align}
\noindent where $\hbC$ and $\hbn$ are determined using the above described inference algorithm. The approximation in Eq.~\eqref{eq:par2} is good whenever most of the mass in the integral in Eq.~\eqref{eq:par2} is around the MAP sequence, $\hbC$.\footnote{Eq.~\eqref{eq:par2} may be considered a first-order Laplace approximation \cite{KassRaftery95}.}  The argument from the right-hand-side of Eq.~\eqref{eq:par2} may be expanded: 
\begin{align} \label{eq:par3}
\log P[\bF| \hbC; \bth] + \log P[\hbn | \bth] &= \sum_{t=1}^T \log P[F_t | \hC_t; \alpha, \beta,\sig] + \sum_{t=1}^T \log P[\hn_t | \lam].
\end{align}
\noindent Note that the two terms in the right-hand-side of Eq.~\eqref{eq:par3} may be optimized separately.  The maximum likelihood estimate (MLE) for the observation parameters, $\{\alpha,\beta,\sig\}$, is therefore given by:
\begin{align} \label{eq:beta,sig}
	\{\halpha, \hbeta,\hsig\} &=  \argmax_{\alpha, \beta,\sig >0} \sum_{t=1}^T \log P[F_t | \hC_t; \beta,\sig]
	=  \argmax_{\beta,\sig >0} 	-\frac{1}{2} (2\pi \sig^2) - \frac{1}{2} \left(\frac{F_t-\alpha(\hC_t+\beta)}{\sig}\right)^2. %
\end{align}
Note that a rescaling of $\alpha$ may be offset by a complementary rescaling of $\bC$ and $\beta$.  Therefore, because the scale of $\bC$ is arbitrary, $\alpha$ can be set to one without loss of generality.  
Plugging $\alpha=1$ into Eq.~\eqref{eq:beta,sig}, and maximizing with respect to $\beta$ yields:
\begin{align}
\hbeta = \argmax_{\beta>0} \sum_{t=1}^T -(F_t - (\hC_t + \beta))^2.
\end{align}
\noindent Computing the gradient with respect to $\beta$, setting the answer to zero, and solving for $\hbeta$, yields $\hbeta = \frac{1}{T} \sum_t (F_t-\hC_t)$.  Similarly, computing the gradient of Eq.~\eqref{eq:beta,sig} with respect to $\sig$, setting it to zero, and solving for $\hsig$ yields:
\begin{align}
\hsig &= \sqrt{\frac{1}{T} \sum_t (F_t - \hC_t - \hbeta)^2},
\end{align}
which is simply the root-mean-square of the residual error.  Finally, the MLE of $\hlam$ is given by solving:
\begin{align}
\hlam &= \argmax_{\lam>0} \sum_t (\log (\lam \Del) - \hn_t \lam \Del),
\end{align}
which, again, computing the gradient with respect to $\lam$, setting it to zero, and solving for $\hlam$, yields $\hlam=T/ (\Del \sum_t \hn_t)$, which is the inverse of the inferred average firing rate.


Iterations stop whenever (i) the iteration number exceeds some upper bound, or (ii) the relative change in likelihood does not exceed some lower bound.  In practice, parameters tend to converge after several iterations, given the above initializations.

\subsection{Spatial filtering} \label{sec:methods:spatial}

In the above, we assumed that the raw movie of fluorescence measurements collected by the experimenter had undergone two stages of preprocessing before filtering.  First, the movie was segmented, to determine regions-of-interest (ROIs), yielding a vector, $\vF_t=(F_{1,t}, \ldots, F_{N_p,t})$, which corresponded to the fluorescence intensity at time $t$ for each of the $N_p$ pixels in the ROI.  Second, at each time $t$, that vector was projected into a scalar, yielding $F_t$, the assumed input to the filter.  In this section, the optimal projection is determined by considering a more general model:
\begin{align} \label{eq:bF}
F_{x,t} &= \alpha_x (C_t + \beta) +  \sig \varepsilon_{x,t}, \qquad &\varepsilon_{x,t} \overset{iid}{\sim} \mathcal{N}(0,1)   
\end{align}
\noindent where $\alpha_x$ scales each pixel, from which some number of photons are contributed due to calcium fluctuations, $C_t$, and others due to baseline fluorescence, $\beta$.  Further, the noise is assumed to be both spatially and temporally white, with standard deviation, $\sig$, in each pixel (this assumption can easily be relaxed, by modifying the covariance matrix of $\varepsilon_{x,t}$).  Performing inference in this more general model proceeds in a  nearly identical manner as before. In particular, the maximization, gradient, and Hessian become:
\begin{align} 
\hbC_{\zzz} 
&= \az  -\frac{1}{2 \sig^2} \norm{\vec{\bF} - \valpha (\bC\T +\bbeta\T)}_F^2 - (\bM \bC )\T \blam  + \zzz \log(\bM \bC)\T\ve{1},  \label{eq:eta4}\\
\ve{g} &= \frac{\valpha}{\sig^2}(\vbF -\valpha({\bC\T} + \bbeta)) - \ve{M}\T\blam + \zzz \ve{M}\T (\ve{M} \bC)^{-1} \label{eq:g2} \\
\ve{H} &= -\frac{\valpha\T \valpha}{\sig^2} \ve{I} - \zzz \ve{M}\T (\ve{M} \bC)^{-2} \ve{M} \label{eq:H2}
\end{align}
\noindent where $\vbF$ is an $N_p \times T$ element matrix, $\valpha$ is a column vector of length $N_p$,  $\bbeta=\beta \ve{1}_T$, where $\ve{1}_T$ is a column vector of ones with length $T$, $\bI$ is an $N_p \times N_p$ identity matrix, and $\norm{x}_F$ indicates the Frobenius norm, i.e.\ $\norm{x}_F^2= \sum_{i,j} x_{i,j}^2$.  Note that to speed up computation, one can first project the $N_c \times T$ dimensional movie onto the spatial filter, $\valpha$, yielding a one-dimensional time series, $\bF$, reducing the problem to evaluating a $T \times 1$ vector norm, as in Eq.~\eqref{eq:eta3}.

Typically, the parameters  $\valpha$ and $\beta$ are unknown, and therefore must be estimated from the data.  Following the strategy developed in the last section, we first initialize the parameters.  Let the initial spatial filter be the median image frame, i.e., $\halpha_x=$ median$_t(F_{x,t})$, and the initial offset be the total movie median, $\hbeta=$ median$_{x,t}(F_{x,t})$.  Given these robust initializations, the maximum likelihood estimator for each $\alpha_x$ is given by:
\begin{subequations} \label{eq:valpha_x}
\begin{align}
\halpha_x &= \argmax_{\alpha_x} P[\bF_x | \hbC] = \argmax_{\alpha_x} \sum_t \log P[F_{x,t} | \hC_t] \\
&=\argmax_{\alpha_x} \sum_t  \left\{-\frac{1}{2} (2\pi \sig^2) - \frac{1}{2\sig^2}\left(F_{x,t} - \alpha_x (\hC_t + \hbeta)\right)^2 \right\} 
= \argmax_{\alpha_x} \sum_t - (F_{x,t} - \alpha_x(\hC_t - \hbeta))^2, \label{eq:alpha_x}
\end{align}
\end{subequations}
which is solved by regressing $(\bC + \hbbeta)$ onto $\bF_x$.  In other words, by computing the gradient of Eq.~\eqref{eq:alpha_x} with respect to $\alpha_x$ and setting to zero, one obtains (using Matlab notation): $\halpha_x = (\bC + \hbbeta)\backslash \bF_x$. 
Computing the gradient of Eq.~\eqref{eq:valpha_x} with respect to $\hbeta$, setting the result to zero, and solving for $\hbeta$, yields:
\begin{align} \label{eq:beta}
	\hbeta = \frac{1}{T N_p} \sum_{t=1}^T \sum_{x=1}^{N_p} \frac{F_{x,t} + \halpha_x  \hC_t}{\halpha_x^2}.
\end{align} 
Iterating these two steps results in a coordinate ascent approach to estimate $\valpha$ and $\beta$ \cite{CONV04}. As in the scalar $F_t$ case, we iterate estimating the parameters of this model, $\bth = \{\valpha, \beta, \sig, \gam, \lam\}$, and the spike train, $\bn$.  Because of the free scale term discussed in section \ref{sec:learn}, the absolute magnitude of $\valpha$ is not identifiable.  Thus, convergence is defined here by the ``shape'' of the spike train converging, i.e., the norm of the difference between the inferred spike trains from subsequent iterations, both normalized such that $\max(\hn_t)=1$.  In practice, this procedure converged after several iterations.

\subsection{Overlapping spatial filters} \label{sec:methods:overlapping}

It is not always possible to segment the movie into pixels containing only fluorescence from a single neuron.  Therefore, the above model can be generalized to incorporate multiple neurons within an ROI.   Specifically, letting the superscript $i$ index the $N_c$ neurons in this ROI yields:  
\begin{align} \label{eq:overlapping}
\vF_t &= \sum_{i=1}^{N_c}\valpha^i (C^i_t + \beta^i) +  \sig\vec{\varepsilon}_t, \qquad &\vec{\varepsilon}_t \overset{iid}{\sim} \mathcal{N}(\ve{0},\bI)   \\
C^i_t &= \gam^i C^i_{t-1} + n^i_t, & n^i_t \overset{iid}{\sim} \text{Poisson}(n^i_t; \lam_i \Del)
\end{align}
\noindent where each neuron is implicitly assumed to be independent, and each pixel is conditionally independent and identically distributed with standard deviation $\sig$, given the underlying calcium signals.  To perform inference in this more general model, let $\ve{1}$ and $\ve{0}$ correspond to an $N_c \times 1$ row vector of ones, and zeros, respectively, and $\bn_t=[n^1_t, \ldots, n^{N_c}_t]$, $\bC_t=[C^1_t, \ldots, C^{N_c}_t]$, and $\bGam=[-\gam^1, \ldots, -\gam^{N_c}]\T$, yielding:
\begin{align} \label{eq:M2}
\ve{M} \bC = 
\begin{bmatrix}
-\bGam & \ve{1} & \ve{0} & \ve{0} & \cdots & \ve{0} \\
\ve{0} & -\bGam & \ve{1} & \ve{0} & \cdots  & \ve{0} \\
\vdots & \ddots & \ddots & \ddots & \ddots & \vdots  \\
\ve{0} & \cdots & \ve{0}  & -\bGam & \ve{1} & \ve{0} \\
\ve{0} & \cdots & \ve{0} & \ve{0} & -\bGam & \ve{1}
\end{bmatrix}
\begin{bmatrix}
\bC_1 \\ \bC_2  \\  \vdots \\ \bC_{T-1} \\ \bC_T  
\end{bmatrix}
= 
\begin{bmatrix}
\bn_1 \\ \bn_2 \\ \vdots \\ \bn_{T-1}
\end{bmatrix}
,
\end{align}
\noindent and proceed as above
.  Note that Eq.~\eqref{eq:M2} is very similar to Eq.~\eqref{eq:M}, except that $\bM$ is no longer tridiagonal, but rather, block tridiagonal (and $\bC_t$ and $\bn_t$ are vectors instead of scalars).  Importantly, the Thomas algorithm, which is a simplified form of Gaussian elimination, finds the solution to linear equations with block tridiagonal matrices in linear time, so the efficiency gained from utilizing the tridiagonal structure above is maintained for this block tridiagonal structure \cite{Press92}.   

If the parameters are unknown, they must be estimated. Define $\balpha_x=[\alpha_x^1, \ldots, \alpha_x^{N_c}]\T$ and $\bbeta=[\beta^1, \ldots, \beta^{N_c}]\T$.  To initialize, let $\bbeta=$ median$_{x,t}(F_{x,t})\ve{1}_{N_p}$, where $\ve{1}_{N_p}$ corresponds to a column vector of length $N_p$. To initialize $\vbalpha=[\valpha_1, \ldots, \valpha_{N_c}]$, the goal is to be able to represent the matrix, $\vbF$, as the sum of only $N_c$ time-varying components, also known as a \emph{low-rank} approximation.  Singular value decomposition is a tool known to find the low-rank approximation to a matrix, with the smallest mean-square-error of all possible low-rank approximations \cite{HornJohnson90}.  Because the ``singular values'' are equivalent to the ``principal components'' of the covariance of the movie, a natural initial estimate for the $N_c$ $\valpha$ vectors are the $N_c$ first principal components.  While other methods to initialize the spatial filters (such as ``independent component analysis'' \cite{MukamelSchnitzer09}) could also work, because fast algorithms for computing the first few principal components are readily available \cite{RokhlinTygert09}, PCA was both sufficiently effective and efficient.  Given these initializations, estimating $\valpha$ and $\bbeta$ follows very similarly as in Eqs.~\eqref{eq:valpha_x} and \eqref{eq:beta}:
\begin{align} \label{eq:valpha}
	\hbalpha_x &= \argmin_{\balpha_x} \sum_{t=1}^T  \left(F_{x,t}  \sum_{i=1}^{N_c} \alpha_x^i (\hC_t^i + \hbeta^i)\right)^2 \\
	\hbeta^i &= \frac{1}{T N_p} \sum_{t=1}^T \sum_{x=1}^{N_p} \frac{F_{x,t} + \halpha_x^i  \hC_t^i}{\halpha_x^2}, \label{eq:vbeta}
\end{align}
where Eq.~\eqref{eq:valpha} is solved efficiently in Matlab using $\balpha_x = (\hbC + \widetilde{\bbeta})\backslash \bF_x$, where $\widetilde{\bbeta}$ is the set of estimated baseline parameters, $\widehat{\bbeta}$, concatenated $T$ times.  Convergence of parameters and spike trains in this model behaved similarly to the scenario described in section \ref{sec:methods:spatial}, assuming the spikes were sufficiently uncorrelated and observations had a sufficiently high SNR.

\subsection{Experimental Methods} \label{sec:exp}

\subsubsection{Slice Preparation and Imaging} 

All animal handling and experimentation was done according to the National Institutes of Health and local Institutional Animal Care and Use Committee guidelines. Somatosensory thalamocortical or coronal slices 350-400 $\mu$m thick were prepared from C57BL/6 mice at age P14 as described \cite{MacLeanYuste05}. Neurons were filled with 50 $\mu$M Oregon Green Bapta 1 hexapotassium salt (OGB-1; Invitrogen, Carlsbad, CA) through the recording pipette or bulk loaded with Fura-2 AM (Invitrogen, Carlsbad, CA). Pipette solution contained 130 mM K-methylsulfate, 2 mM MgCl$_2$, $0.6$ mM EGTA, 10 mM HEPES, 4 mM ATP-Mg, and $0.3$ mM GTP-Tris, pH 7.2 (295 mOsm).  After cells were fully loaded with dye, imaging was done by using a modified BX50-WI upright microscope (Olympus, Melville, NY).  Image acquisition was performed with the C9100-12 CCD camera from Hamamatsu Photonics (Shizuoka, Japan) with arclamp illumination with excitation and emission bandpass filters at 480-500 nm and 510-550 nm, respectively  (Chroma, Rockingham, VT) for Oregon Green. Imaging of Fura-2 loaded slices was performed with a confocal spinning disk (Solamere Technology Group, Salt Lake City, UT) and an Orca CCD camera from Hamamatsu Photonics (Shizuoka, Japan). Images were saved and analyzed using custom software written in Matlab (Mathworks, Natick, MA).

\subsubsection{Electrophysiology}

All recordings were made using the Multiclamp 700B amplifier (Molecular Devices, Sunnyvale, CA), digitized with National Instruments 6259 multichannel cards and recorded using custom software written using the LabView platform (National Instruments, Austin, TX) .  Square pulses of sufficient amplitude to yield the desired number of action potentials were given as current commands to the amplifier using the LabView and National Instruments system.

\subsubsection{Fluorescence preprocessing}

Traces were extracted using custom Matlab scripts to (i) segment the mean image into ROIs, and then (ii) average all the pixels within the ROI.  The Fura-2 fluorescence traces were inverted.  As some slow drift was sometimes present in the traces, each trace was Fourier transformed, and all frequencies below $0.5$ Hz were set to zero ($0.5$ Hz was chosen by eye), and the resulting fluorescence trace was then normalized to be between zero and one.

\section{Results} \label{sec:results}

\subsection{Main Result} \label{sec:main}

\begin{figure}[h!]
\centering \includegraphics[width=.9\linewidth]{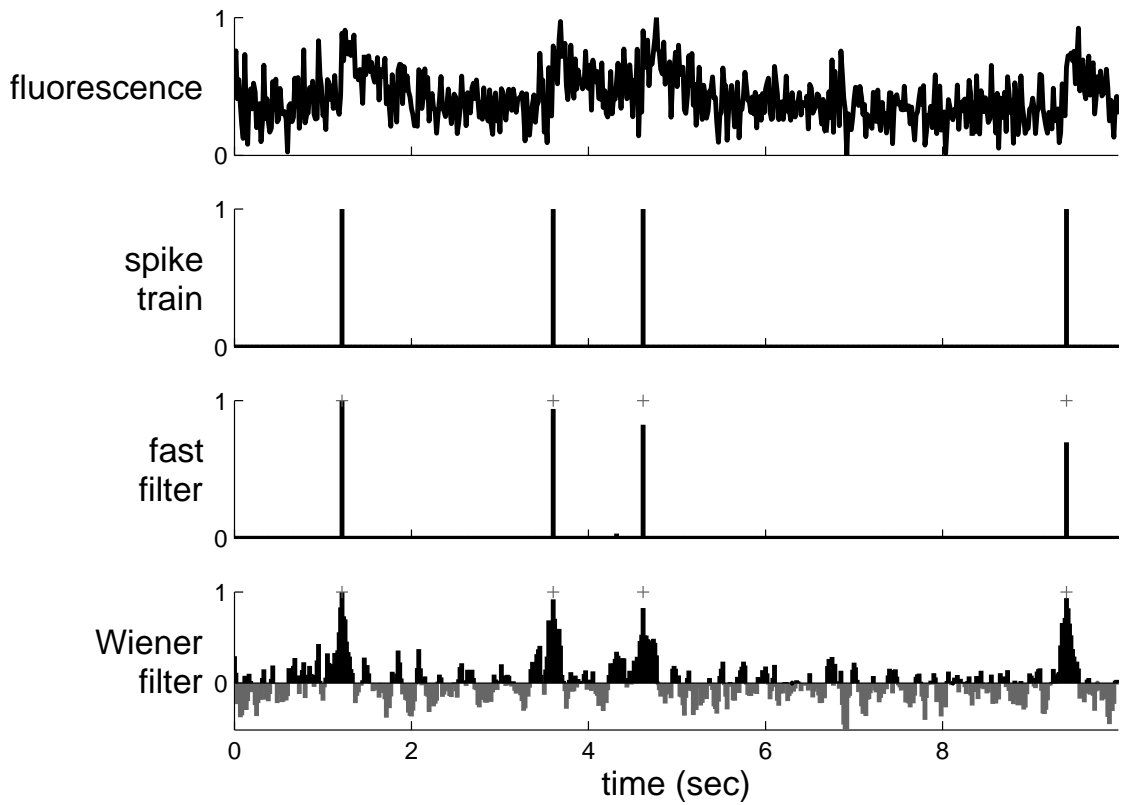}
\caption[\foopsi filter outperforms Wiener filter]{The \foopsi filter's inferred spike train is significantly more accurate than the output of the optimal linear deconvolution (Wiener filter) on typical simulated data. Note that neither filter constrains the inference to be a sequence of integers; rather, the \foopsi filter relaxes the constraint to allow all non-negative numbers, and the Wiener filter allows for all real numbers.  The restriction of the \foopsi filter to exclude negative numbers eliminates the ringing effect seen in the Wiener filter output, resulting in a much cleaner inference.  Note that the magnitude of the inferred spikes in the \foopsi filter output is proportional to the inferred calcium jump size.  Top panel: fluorescence trace.  Second panel: spike train.  Third panel: \foopsi filter inference.  Bottom panel: Wiener filter inference.  Note that the gray bars in the bottom panel indicate \emph{negative} spikes. Black '$+$'s in bottom two panels indicate true spike times.  Simulation details: $T\approx 3000$ time steps, $\Del=5$ msec, $\alpha=1$, $\beta=0$, $\sig=0.3$, $\tau=1$ sec, $\lam=1$ Hz. Conventions for other figures as above, unless otherwise indicated.} \label{fig:woopsi_inf}
\end{figure}

The main result of this paper is that the \foopsi filter can find the approximately most likely spike train, $\hbn$, very efficiently, and that this approach yields more accurate spike train estimates than optimal linear deconvolution.  Fig. \ref{fig:woopsi_inf} depicts a simulation showing this result. Clearly, the \foopsi filter's inferred ``spike train'' (third panel) more closely resembles the true spike train (second panel) than the optimal linear deconvolution's inferred spike train (bottom panel; Wiener filter).  Note that neither filter results in an integer sequence, but rather, each infers a real number at each time.

The Wiener filter implicitly approximates the Poisson spike rate with a Gaussian spike rate (see Appendix \ref{sec:wiener} for details).  A Poisson spike rate indicates that in each frame, the number of possible spikes is an integer, 0, 1, 2, \ldots.  The Gaussian approximation, however, allows for any real number of spikes in each frame, including both partial spikes, e.g., 1.4, and \emph{negative} spikes, e.g., -0.8.  While a Gaussian well approximates a Poisson distribution when rates are about 10 spikes per frame, this example is very far from that regime, so the Gaussian approximation performs relatively poorly.  More specifically, the Wiener filter exhibits a ``ringing'' effect.  Whenever fluorescence drops rapidly, the most likely underlying signal is a proportional drop.  Because the Wiener filter does not impose a non-negative constraint on the underlying signal (which, in this case, is a spike train), it infers such a drop.  After such a drop has been inferred, since no corresponding drop occurred in the true underlying signal here, a complementary jump is often then inferred, to ``re-align'' the inferred signal with the observations.  This oscillatory behavior results in poor inference quality. The non-negative constraint imposed by the \foopsi filter prevents this because the underlying signal never drops below zero, so the complementary jump never occurs either.


The inferred ``spikes'', however, are still not binary events when using the \foopsi filter.  This is a by-product of approximating the Poisson distribution on spikes with an exponential (cf.~ Eq.~\eqref{eq:obj2}), because the exponential is a \emph{continuous} distribution, versus the Poisson, which is discrete.  The height of each spike is therefore proportional to the inferred calcium jump size, and can be thought of as a proxy for the confidence with which the algorithm believes a spike occurred.   Importantly, by utilizing the Gaussian elimination and interior-point methods, as described in the Methods section, the computational complexity of \foopsi filter is the same as an efficient implementation of the Wiener filter.

\begin{figure}[h!]
\centering \includegraphics[width=.9\linewidth]{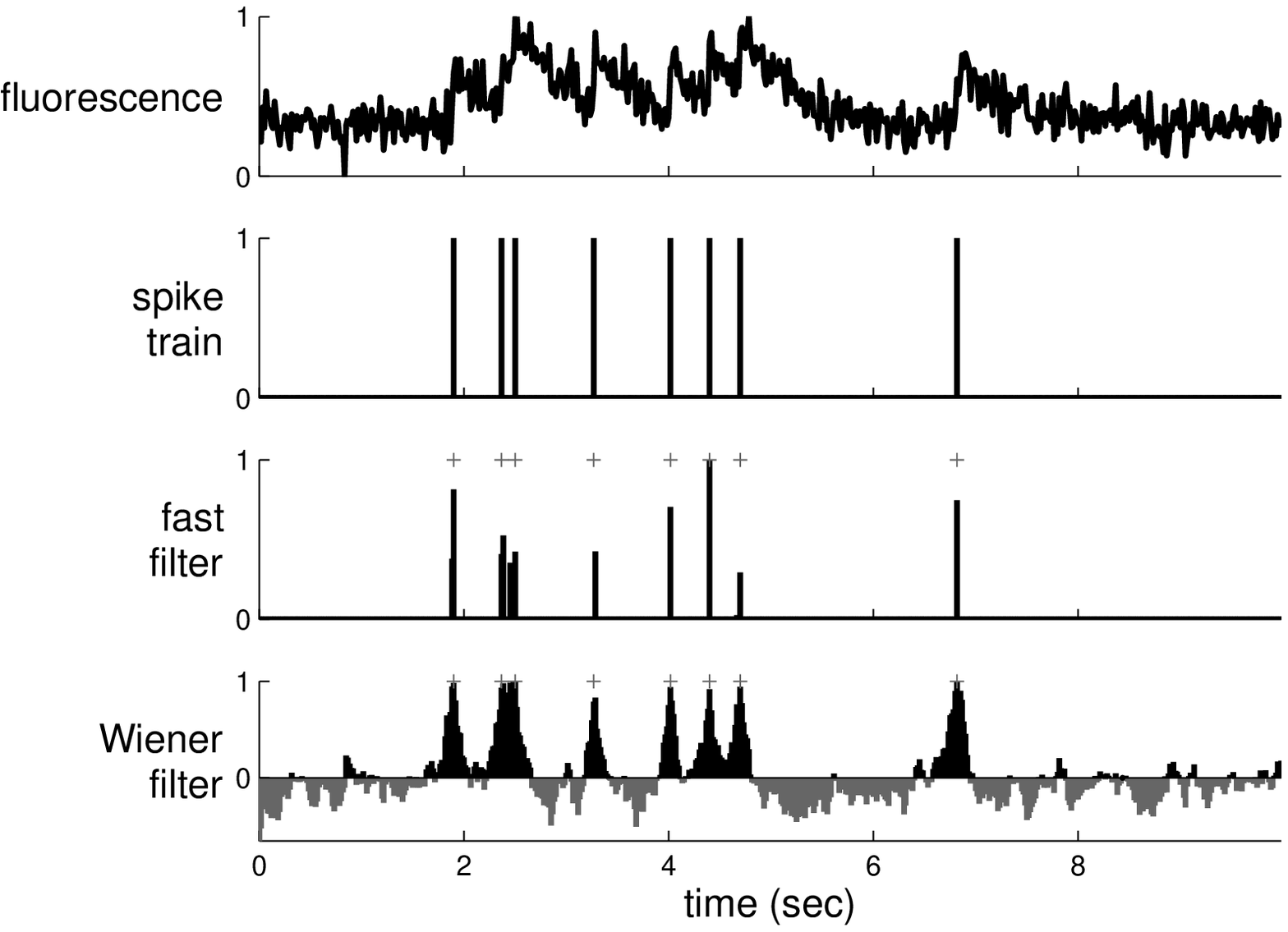}
\caption[parameters may be estimated using the \foopsi filter]{The \foopsi filter significantly outperforms the Wiener filter, even when the parameters unknown.  For both filters, the appropriate parameters were estimated using only the data shown above, unlike Figure \ref{fig:woopsi_inf}, in which the true parameters were provided to the filters. Simulation details as in Figure \ref{fig:woopsi_inf}.} \label{fig:woopsi_learn}
\end{figure}

Although in Figure \ref{fig:woopsi_inf} the model parameters were provided, in the general case, the parameters are unknown, and must therefore be estimated from the observations (as described in section \ref{sec:learn}). Importantly, this algorithm does not require ``training'' data, i.e., there is no need for joint imaging and electrophyiological experiments to estimate the parameters governing the relationship between the two.  Figure \ref{fig:woopsi_learn} shows another simulated example; in this example, however, the parameters are estimated from the observed fluorescence trace alone.  Again, it is clear that the \foopsi filter far outperforms the Wiener filter.

\begin{figure}[h!]
\centering \includegraphics[width=.9\linewidth]{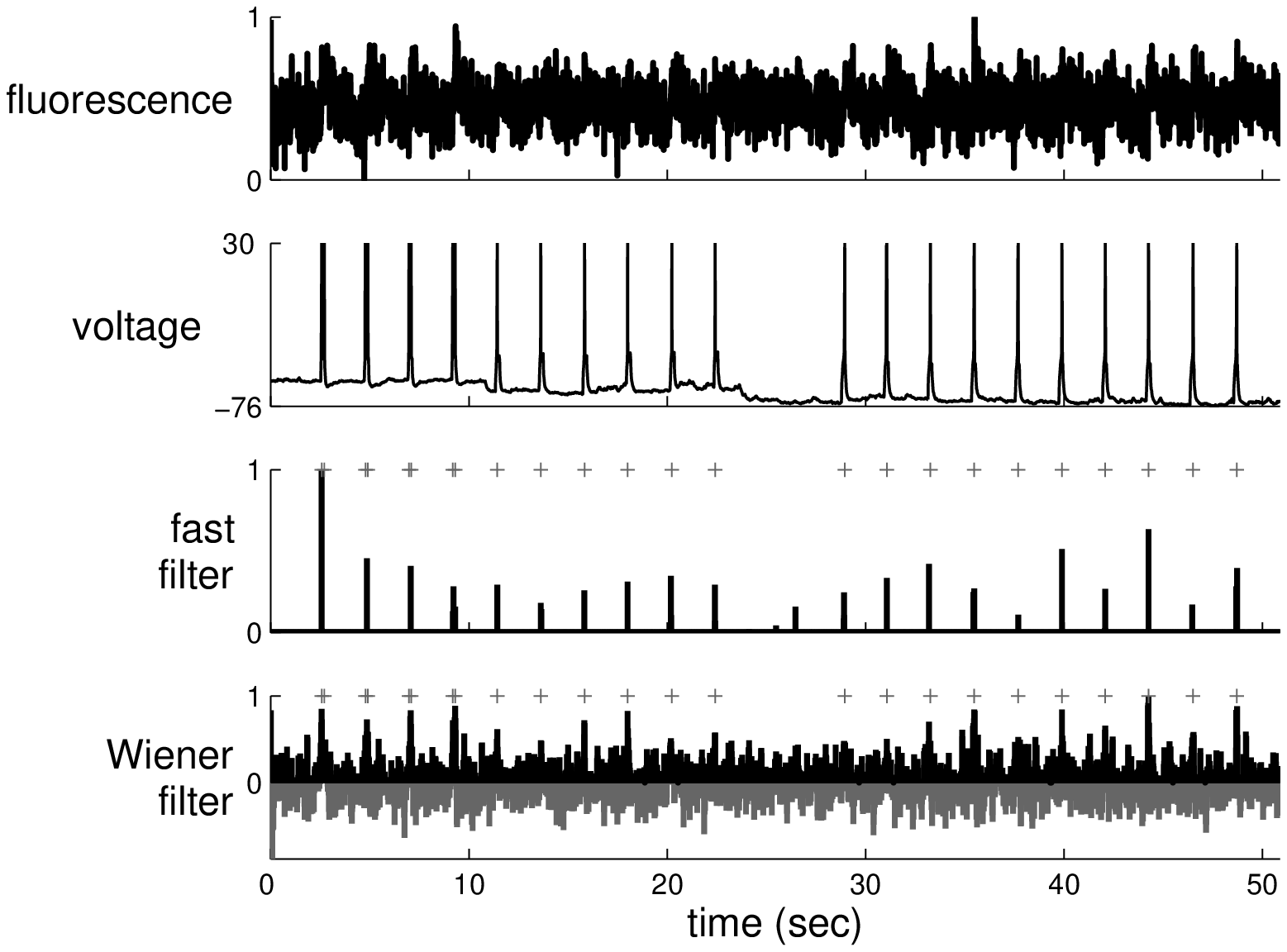}
\caption[\foopsi filter outperforms Wiener filter on biological data]{The \foopsi filter significantly outperforms the Wiener filter on typical \emph{in vitro} data, using OGB-1. Note that all the parameters for both filters were estimated only from the fluorescence data in the top panel (ie, not considering the voltage data at all).  Again, '$+$'s denote true spike times extracted from the patch data, not inferred spike times from $\bF$.} \label{fig:woopsi_data}
\end{figure}

Given the above two results, the \foopsi filter was applied to real data.  More specifically, by jointly recording electrophysiologically and imaging, the true spike times are known, and the accuracy of the two filters can be compared.  Figure \ref{fig:woopsi_data} shows a result typical of the 12 joint electrophysiological and imaging experiments conducted. Although it is difficult to see in this figure, the first four ``events'' are actually pairs of spikes, which is reflected by the width and height of the corresponding inferred spikes when using the \foopsi filter. 
This suggests that although the scale of $\bn$ is arbitrary, the \foopsi filter can correctly ascertain the number of spikes within spike events.  

\begin{figure}[h!]
\centering \includegraphics[width=.9\linewidth]{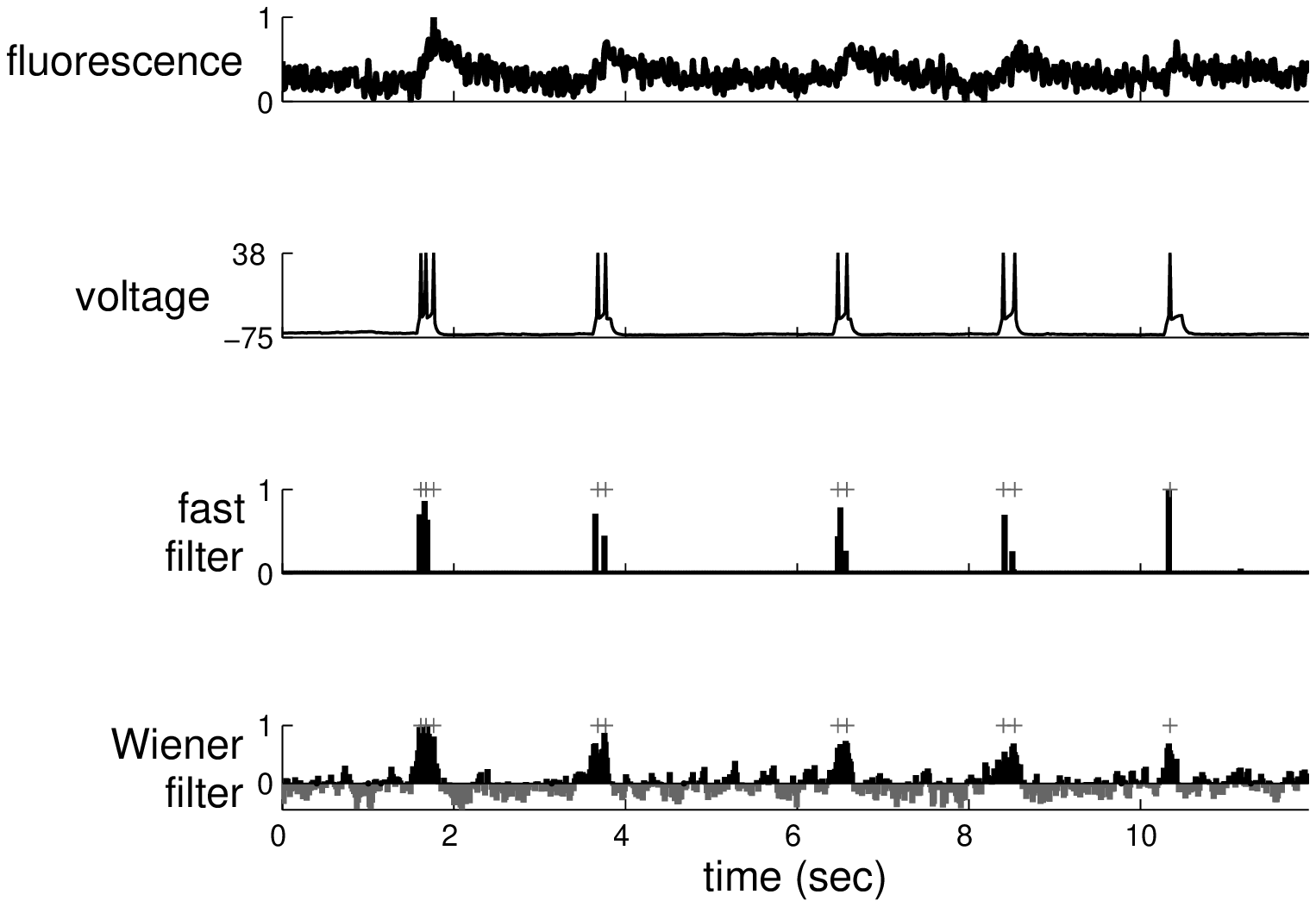}
\caption[\foopsi filter outperforms Wiener filter on multi-spike events]{The \foopsi filter can often resolve the correct number of spikes within each spiking event, while imaging using OGB-1, given sufficiently high SNR.  It is difficult, if not impossible, to count the number of spikes given the Wiener filter output.  Recording and fitting parameters as in Figure \ref{fig:woopsi_data}. Note that the parameters were estimated using a 60 sec long recording, of which only a fraction is shown here, to more clearly depict the number of spikes per event.  } \label{fig:woopsi_data_doublets}
\end{figure}

Figure \ref{fig:woopsi_data_doublets} further evaluates this claim.  While recording and imaging, the cell was forced to spike once, twice, or thrice, for each spiking event.  Na\"{i}vely, this would suggest that an algorithm based on a purely linear model would struggle to resolve spike fidelity in this high frequency spiking regime.  However, the \foopsi filter infers the correct number of spikes in each event.  On the contrary, there is no obvious way to count the number of spikes within each event when using the Wiener filter.

\subsection{Online analysis of spike trains using the \foopsi filter}

\begin{figure}[h!]
\begin{centering} 
\includegraphics[width=1\linewidth]{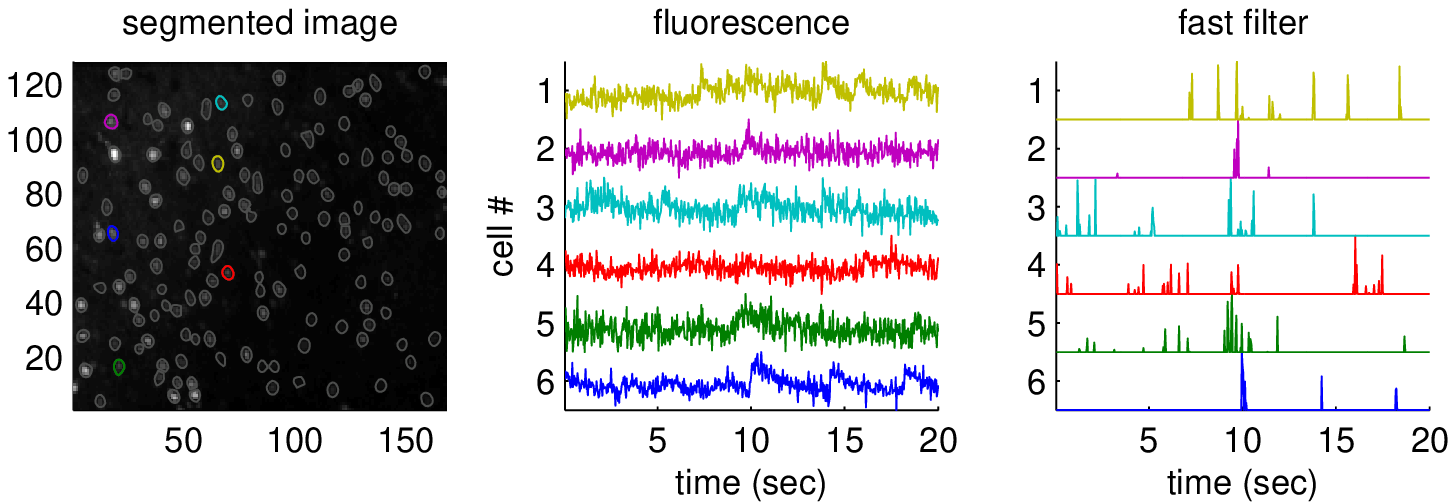}
\end{centering}
\caption[\foopsi filter is robust and works online for populations of neurons]{The \foopsi filter infers spike trains from a large population of neurons imaged simultaneously \emph{in vitro}, using Fura-2, faster than real-time.  Specifically, inferring the spike trains from this 400 sec long movie including 136 neurons requires only about 40 sec on a standard laptop computer. The inferred spike trains much more clearly convey neural activity than the raw fluorescence traces.  Although no intracellular ``ground truth'' is available on this population data, the noise seems to be reduced, consistent with the other examples with ground truth.  Left panel: Mean image field, segmented into ROIs each containing a single neuron.  Middle panel: example fluorescence traces.  Right panel: \foopsi filter output corresponding to each associated trace. Note that neuron identity is indicated by color across the three panels.} \label{fig:pop}
\end{figure}

A central aim for this work was the development of an algorithm that infers spikes fast enough to use online while imaging a large population of neurons (eg, $\approx 100$).  Figure \ref{fig:pop} shows a segment of the results of running the \foopsi filter on 136 neurons, recorded simultaneously, as described in section \ref{sec:exp}.  Note that the filtered fluorescence signals show fluctuations in spiking much more clearly than the unfiltered fluorescence trace. These spike trains were inferred in less than imaging time, meaning that one could infer spike trains for the past experiment while conducting the subsequent experiment. More specifically, a movie with 5,000 frames of 100 neurons can be analyzed in about ten seconds on a standard desktop computer.  Thus, if that movie was recorded at 50 Hz, while collecting the data required 100 seconds, inferring spikes only required ten seconds, a ten-fold improvement over real-time.

\subsection{Extensions}

Section \ref{sec:model} describes a simple principled first-order model relating the spike trains to the fluorescence trace. A number of the simplifying assumptions can be straightforwardly relaxed, as described below.

\subsubsection{Replacing Gaussian observations with Poisson}

In the above, observations were assumed to have a Gaussian distribution.  The statistics of photon emission and counting, however, suggest that a Poisson distribution would be more natural, especially for two-photon data \cite{SjulsonMiesenbock07}, yielding:
\begin{align} \label{eq:poiss}
	F_t \overset{iid}{\sim}\text{Poisson}(\alpha C_t + \beta).
\end{align}
One additional advantage to this model over the Gaussian model, is that the variance parameter, $\sig^2$, no longer exists, which might make learning the parameters simpler.  Importantly, the log-posterior is still concave in $\bC$, as
the prior remains unchanged, and the new log-likelihood term is a sum of terms concave in $\bC$:
\begin{align}
	\log P[\bF | \bC] = \sum_{t=1}^T \log P[F_t | C_t ] = \sum_{t=1}^T \{F_t \log (\alpha C_t + \beta) -(\alpha C_t + \beta) - \log(F_t !)\}.
\end{align}
The gradient and Hessian of the log-posterior can therefore be computed analytically by substituting the above likelihood terms for those implied by Eq.~\eqref{eq:F}.  In practice, however, modifying the filter for this model extension did not seem to significantly improve inference results in any simulations or data (not shown).

\subsubsection{Allowing for a time-varying prior}

In Eq.~\eqref{eq:n}, the rate of spiking is a constant.  Often, additional knowledge about the experiment, including external stimuli, or other neurons spiking, can provide strong time-varying prior information \cite{VogelsteinPaninski09}.  A simple model modification can incorporate that feature:
\begin{align}
	n_t &\overset{iid}{\sim} \text{Poisson}(\lam_t \Del),
\end{align}
where $\lam_t$ is now a function of time.  Approximating this time-varying Poisson with a time-varying exponential with the same time-varying mean (similar to Eq.~\eqref{eq:obj2}), and letting $\blam = [\lam_1, \ldots, \lam_T]\T \Del$, yields an objective function identical to Eq.~\eqref{eq:eta3}, so log-concavity is maintained, and the same techniques may be applied.  However, as above, this model extension did not yield any significantly improved filtering results (not shown).

\subsubsection{Saturating fluorescence}

Although all the above models assumed a \emph{linear} relationship between $F_t$ and $C_t$, the relationship between fluorescence and calcium is typically better approximated by the nonlinear Hill equation \cite{PologrutoSvoboda04}. Modifying Eq.~\eqref{eq:F} to reflect this change yields: 
\begin{align}
	F_t &= \alpha \frac{C_t}{C_t+k_d} + \beta +  \sig \varepsilon_t, \qquad \varepsilon_t \overset{iid}{\sim} \mN(0,1).
\end{align}
Importantly, log-concavity of the posterior is no longer guaranteed in this nonlinear model, meaning that converging to the global maximum is no longer guaranteed.  Assuming a good initialization can be found, however, if this model is more accurate, then ascending the gradient for this model might yield improved inference results.  In practice, initializing with the  inference from the \foopsi filter assuming a linear model (eg, Eq.~\eqref{eq:overlapping}) often resulted in nearly equally accurate inference, but inference assuming the above nonlinearity was far less robust than the inference assuming the linear model (not shown).  

\subsubsection{Using the \foopsi filter to initialize the sequential Monte Carlo filter}

\begin{figure}[h!]
\centering \includegraphics[width=.9\linewidth]{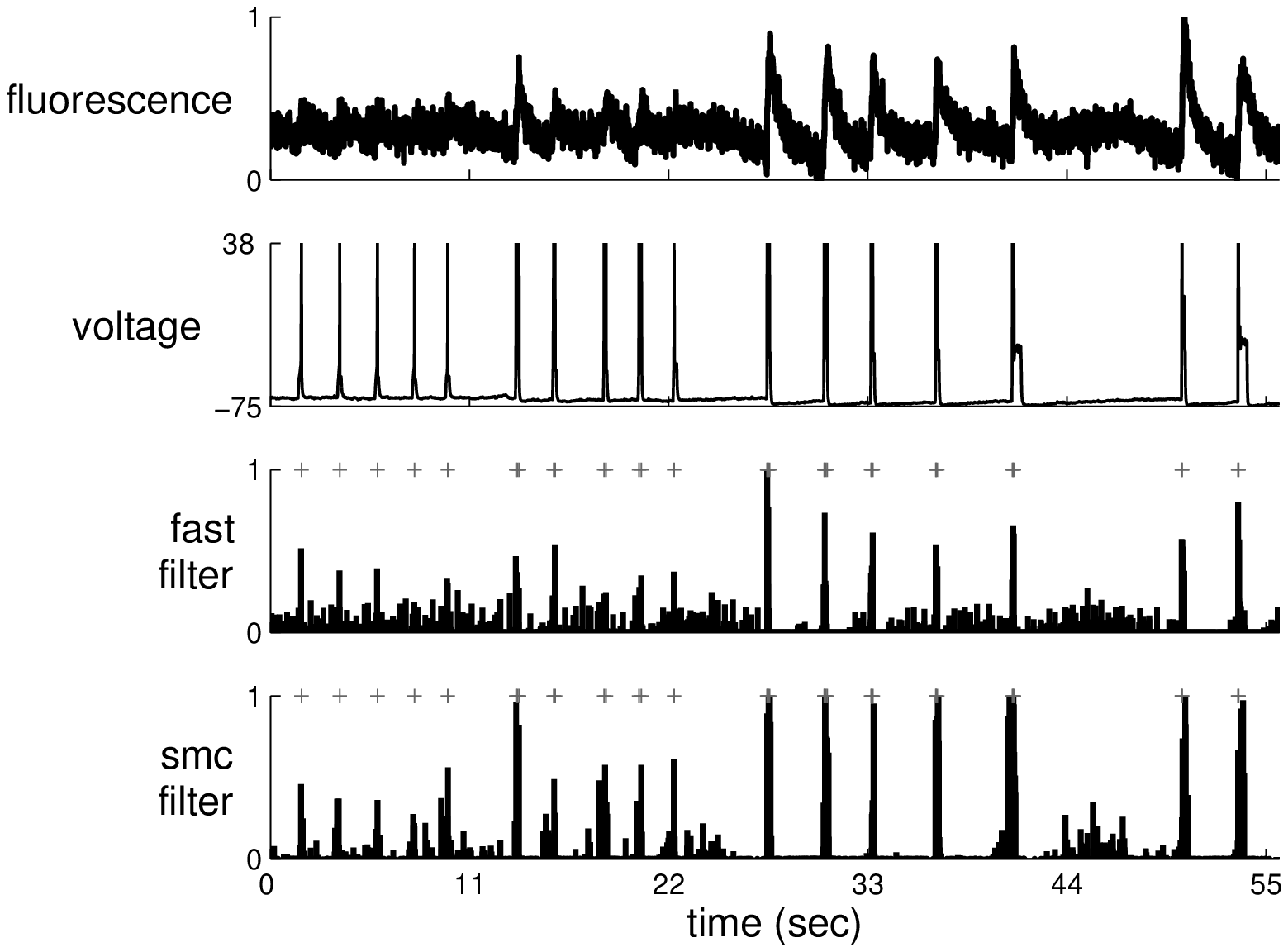}
\caption[\foopsi filter can initialize Wiener filter]{The \foopsi filter effectively initializes the parameters for the SMC filter, significantly reducing the number of expectation-maximization iterations to convergence, on typical \emph{in vitro} data, using OGB-1.  Note that while the \foopsi filter clearly infers the spiking events in the end of the trace, those in the beginning of the trace are less clear.  On the other hand, the SMC filter more clearly separates non-spiking activity from true spikes.  Also note that the ordinate on the bottom panel corresponds to the inferred probability of a spike having occurred in each frame.} \label{fig:smc_init}
\end{figure}

A sequential Monte Carlo (SMC) method to infer spike trains can incorporate this saturating nonlinearity, as well as the other model extensions discussed above \cite{VogelsteinPaninski09} . However, this SMC filter is not nearly as computationally efficient as the fast filter proposed here.  Like the \foopsi filter, the SMC filter estimates the model parameters in a completely unsupervised fashion, i.e.,  from the fluorescence observations, using an expectation-maximization algorithm (which requires iterating between computing the expected value of the hidden variables --- $\bC$ and $\bn$ --- and updating the paramters).  In \cite{VogelsteinPaninski09}, parameters for the SMC filter were initialized based on other data.  While effective, this initialization was often far from the final estimates, and therefore, required a relatively large number of iterations (eg, 20--25) before converging.  Thus, it seemed that the \foopsi filter could be used to obtain an improvement to the initial parameter estimates, given an appropriate rescaling to account for the nonlinearity, thereby reducing the required number of iterations to convergence.  Indeed, Figure \ref{fig:smc_init} shows how the SMC filter outperforms the \foopsi filter on biological data, and only required 3--5 iterations to converge on this data, given the initialization from the \foopsi filter (which was typical).  Note that the first few events of the spike train are individual spikes, resulting in relatively small fluorescence fluctuations, whereas the next events are actually spike doublets or triplets, causing a much larger fluorescence fluctuation.  Only the SMC filter picks up the individual spikes in this trace, a result typical when the effective signal-to-noise ratio (SNR) is poor.  Thus, these two inference algorithms are complementary: the \foopsi filter can be used for rapid, online inference, and for initializing the SMC filter, which can then be used to further refine the spike train estimate.  Importantly, although the SMC filter often outperforms the \foopsi filter, the \foopsi filter is more robust, meaning that it more often works ``out-of-the-box.''

\subsection{Spatial filter} \label{sec:results:spatial}

In the above, the filters operated on one-dimensional fluorescence traces. Typically, the data are time-series of images which are first segmented into regions-of-interest (ROI), and then (usually) averaged to obtain $F_t$.  In theory, one could improve the effective SNR of the fluorescence trace by scaling each pixel relative to one another.  In particular, pixels not containing any information about calcium fluctuations can be ignored, and pixels that are partially anti-correlated with one another could have weights with opposing signs.  

\begin{figure}[h!]
\centering \includegraphics[width=.9\linewidth]{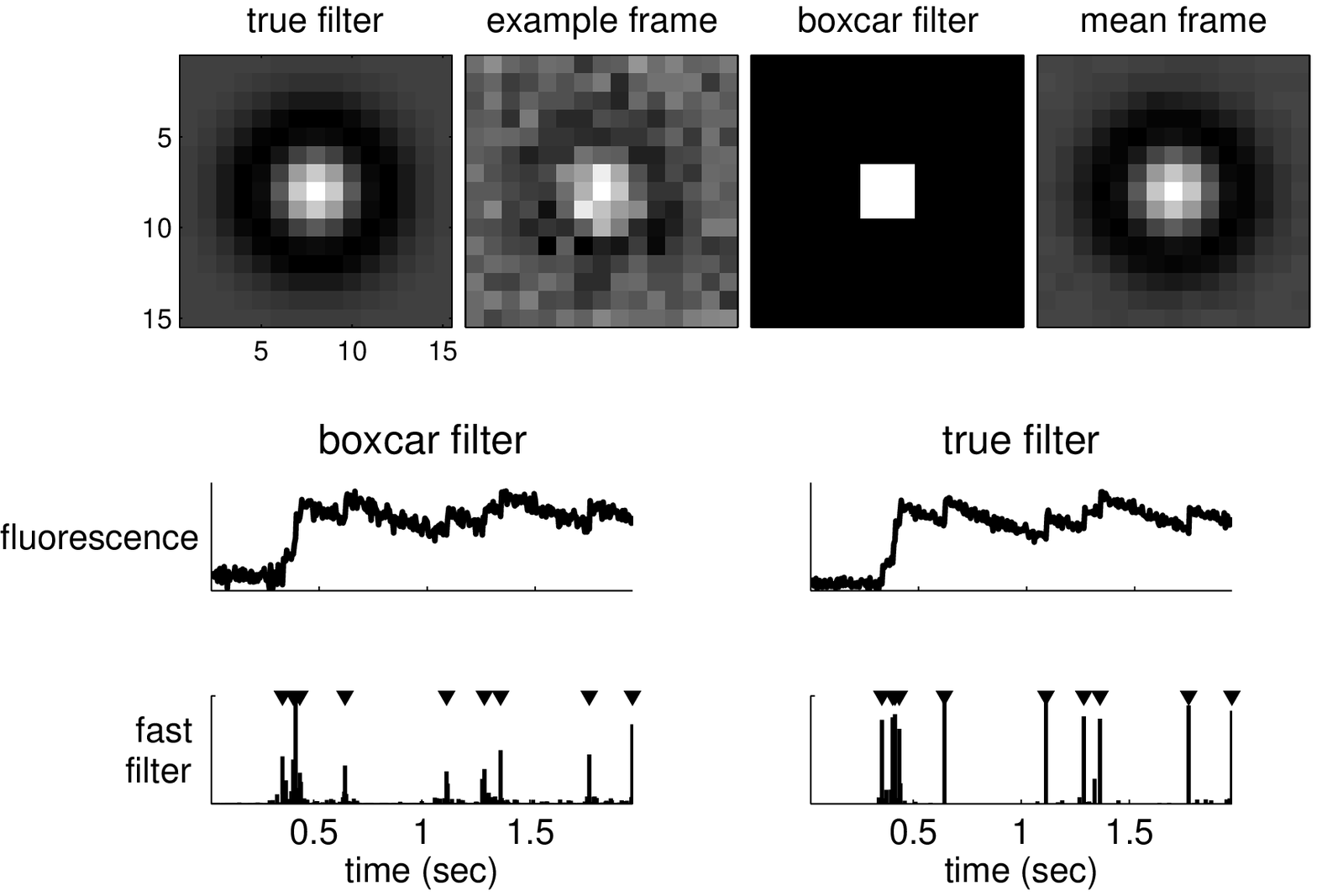}
\caption[spatial filtering can improve effective SNR]{A simulation demonstrating that using a better spatial filter can significantly enhance the effective SNR. The true spatial filter was a difference of Gaussians: a positively weighted Gaussian of small width, and a negatively weighted Gaussian with larger width (both with the same center).  Top row far left: true spatial filter.  Top row second from left: example movie frame. Top row second from right: typical spatial filter.   Top row far right: mean frame.  Middle row left: fluorescence trace using the boxcar spatial filter. Bottom row left: \foopsi filter output using the boxcar spatial filter.  Middle row right: fluorescence trace using true spatial filter.  Bottom right: \foopsi filter output using true spatial filter. Simulation details: $\valpha=\mN(\ve{0},2 \bI)-1.1 \mN(\ve{0},2.5 \bI)$ where $\mN(\ve{\mu},\ve{\Sig})$ indicates a two-dimensional Gaussian with mean $\ve{\mu}$ and covariance matrix $\ve{\Sig}$, $\bbeta=1$, $\tau=0.85$ sec, $\lam=5$ Hz.} \label{fig:spatial} 
\end{figure}

Figure \ref{fig:spatial} demonstrates the potential utility of this approach.  The top row shows different depictions of an ROI containing a single neuron.  On the far left panel is the true spatial filter for this neuron.  This particular spatial filter was chosen based on experience analyzing both \emph{in vitro} and \emph{in vivo} movies; often, it seems that the pixels immediately around the soma are anti-correlated with those in the soma.  This effect is possibly due to the influx of calcium from the extracellular space immediately around the soma.  The simulated movie (not shown) is relatively noisy, as indicated by the second panel, which depicts an exemplary image frame.  The standard approach, given such a noisy movie, would be to first segment the movie to find an ROI corresponding to the soma of this cell, and then spatially average all the pixels found to be within this ROI.  The third panel shows this standard ``boxcar spatial filter''.  The fourth panel shows the mean frame.  Clearly, this mean frame is very similar to the true spatial filter.  

The bottom panels of Figure \ref{fig:spatial} depict the effect of using the true spatial filter, versus the typical one. The left side shows the fluorescence trace and its associated spike inference obtained from using the typical spatial filter.  The right side shows the same when using the true spatial filter.  Clearly, the true spatial filter results in a much cleaner fluorescence trace and spike inference.  When the true spatial filter is a single Gaussian, the boxcar spatial filter works about as well as the true spatial filter (not shown).

\subsection{Overlapping spatial filters} \label{sec:results:overlapping}

\begin{figure}[h!]
\centering \includegraphics[width=.9\linewidth]{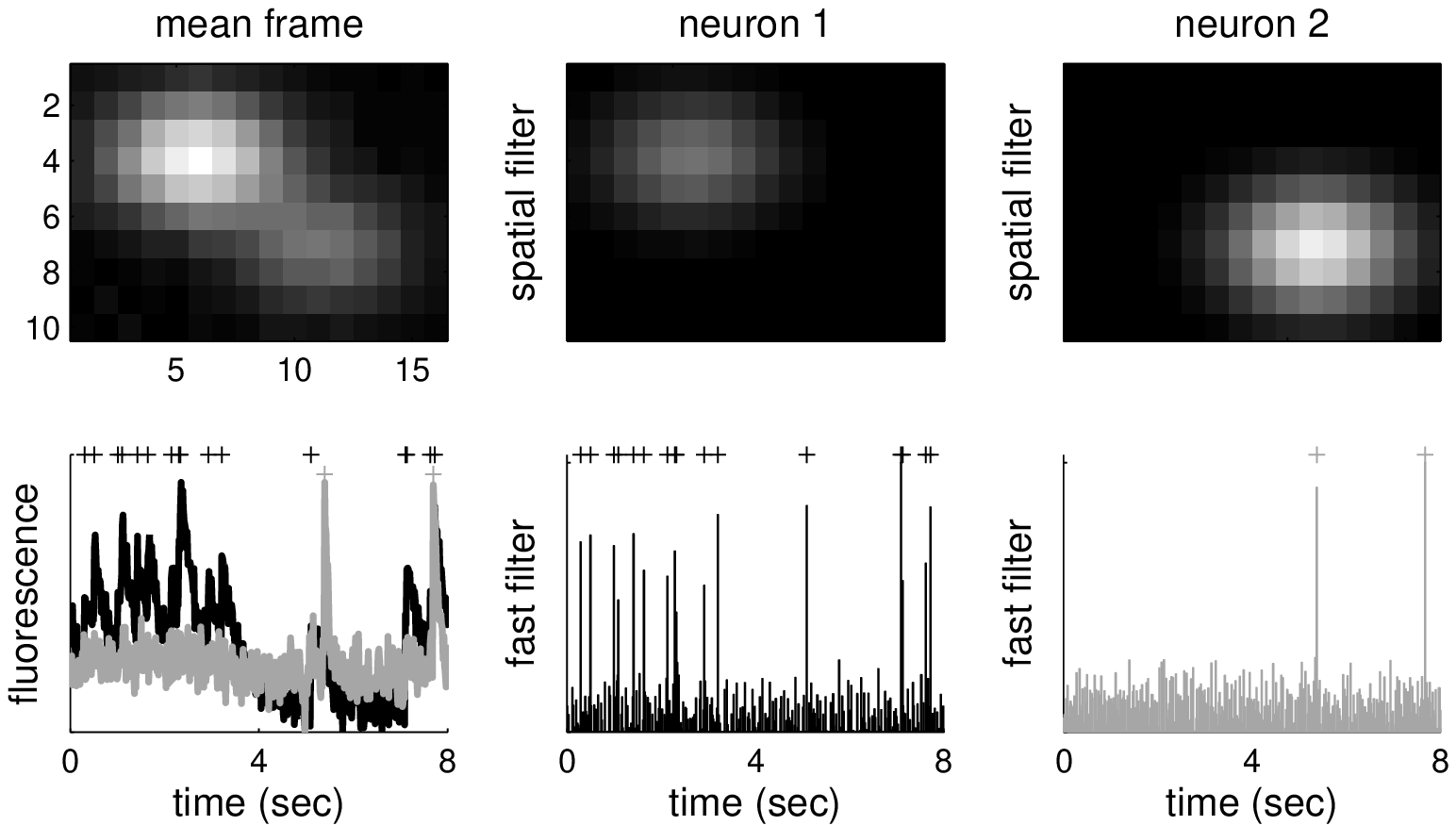}
\caption[overlapping spatial filters are not problematic]{Simulation showing that even when two neurons' spatial filters are overlapping, one can separate the two spike trains by spatial filtering. Top left panel: mean frame from the movie.  Bottom left: optimal one-dimensional fluorescence projections for the neuron 1 (black line) and neuron 2 (gray line), and their respective spike trains (black and gray '$+$' symbols, respectively).  Top middle panel: the true spatial filter for neuron 1.  Bottom middle panel: inferred (black line) and true (black '$+$' symbols) spike trains.  Top right panel: the true spatial filter for neuron 2.   Bottom right panel: inferred (gray line) and true (gray '$+$' symbols) spike trains. Simulation details: $\valpha^1=\mN([-1.8, 1.8]\T,2 \bI)\, \valpha^2=\mN([1.8, -1.8]\T,5 \bI)$, $\bbeta=[1, 1]\T$, $\tau=[0.5, 0.5]\T$ sec, $\lam=[1.5, 1.5]\T$ Hz.} \label{fig:spatial_multi_inf}
\end{figure}

The above shows that if a ROI contains only a single neuron, the effective SNR can be enhanced by spatially filtering.  However, this analysis assumes that only a single neuron is in the ROI.  Often, ROIs are overlapping, or nearly overlapping, making the segmentation problem more difficult.  Therefore, it is desirable to have an ability to crudely segment, yielding only a few neurons in each ROI, and then spatially filter within each ROI to pick out the spike trains from each neuron.  This may be achieved in a principled manner by generalizing the model as described in section \ref{sec:methods:overlapping}.  Figure \ref{fig:spatial_multi_inf} shows how this approach can separate the two signals, assuming that the spatial filters of the two neurons are known.  

\begin{figure}[h!]
\centering \includegraphics[width=.9\linewidth]{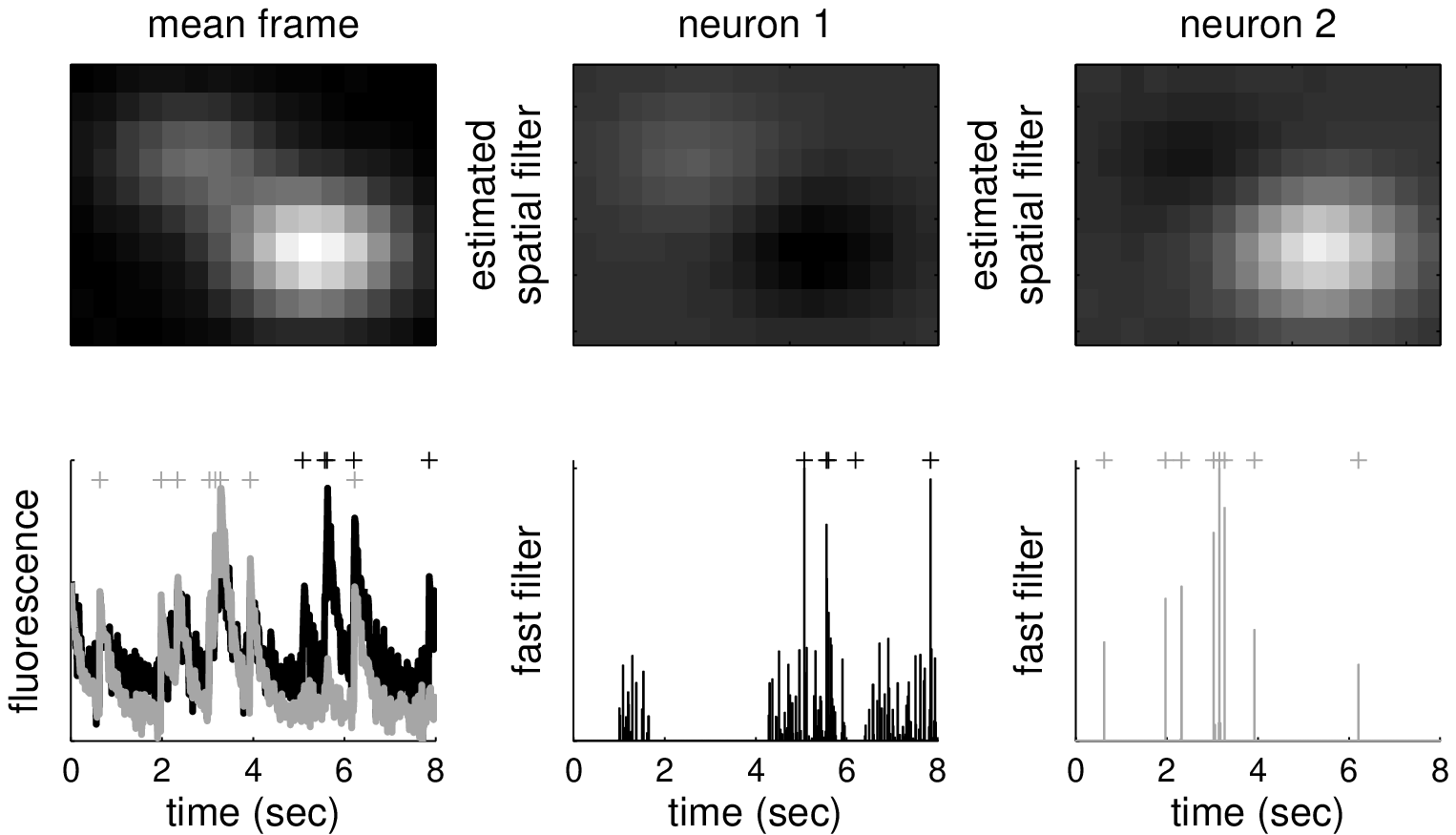}
\caption[overlapping spatial filters can be estimated]{Simulation showing that even when two neurons' spatial filters are largely overlapping, spatial filters that together are linearly related to the true spatial filters can be inferred, to separate the two signals. Simulation details as above. Note that the spatial fields are sufficiently overlapping to cause significant ``bleed-through'' between the two signals.  In particular, this is clear from the rise in the black line in the first few seconds of the bottom left panel, which should be fluorescence due to neuron 1, but is in fact due to spiking from neuron 2. Regardless, the spike trains are accurately inferred here.  Simulation parameters as in Figure \ref{fig:spatial_multi_inf}.} \label{fig:spatial_multi_learn}
\end{figure}

Typically, the true spatial filters of the neurons in the ROI will be unknown, and thus, must be estimated from the data.  This problem may be considered a special case of blind source separation \cite{BellSejnowski95, MukamelSchnitzer09}. Figure \ref{fig:spatial_multi_learn} shows that multiple signals can be separated, with reasonable assumptions on correlations between the signals, and SNR.  Note that separation occurs even though the signal is overlapping in several pixels (top left panel), leading to a ``bleed-through'' effect in the one-dimensional fluorescence projections (bottom left panel). The inferred filters (top middle and right panels) are not the true filters, but rather, their sum is linearly related to the sum of the true filters.  Regardless, the inferred spike trains are well separated (bottom middle and right panels).

\section{Discussion} \label{sec:dis}


This work describes an algorithm that approximates the \emph{maximum a posteriori} (MAP) spike train, given a calcium fluorescence movie.  The approximation is required because finding the actual MAP estimate is not currently computationally tractable.  Replacing the assumed Poisson distribution on spikes with an exponential distribution yields a log-concave optimization problem, which can be solved using standard gradient ascent techniques (such as Newton-Raphson).  This exponential distribution has an advantage over a Gaussian distribution by restricting spikes to be positive, which improves inference quality (cf. Figure \ref{fig:woopsi_inf}), and is a better approximation to a Poisson distribution with low rate.  
Furthermore, by utilizing the special banded structure of the Hessian matrix of the log-posterior, this approximate MAP spike train can be inferred fast enough on standard computers to use it for online analyses.  Finally, all the parameters can be estimated from only the fluorescence observations, obviating the need for joint electrophysiology and imaging (cf. Figure \ref{fig:woopsi_learn}).  This approach is robust, in that it works ``out-of-the-box'' on all the \emph{in vivo} and \emph{in vitro} data analyzed (cf. Figure \ref{fig:woopsi_data}).

Ideally, one could compute the full joint posterior of entire spike trains, conditioned on the fluorescence data.  This distribution is analytically intractable, due to the Poisson assumption on spike trains.  A Bayesian approach could use Markov Chain Monte Carlo methods to recursively sample spikes until a whole sample spike train is obtained \cite{AndrieuDoucet01,MishchenkoPaninski09}.  Because a central aim here was computational expediency, a ``greedy'' approach is natural: i.e.,  recursively sample the most likely spike, update the posterior, and repeat until the posterior stops increasing.  Template matching, projection pursuit regression \cite{FS81}, and matching pursuit \cite{MallatZhang93} are examples of such a greedy approach (Greenberg et al's algorithm \cite{GreenbergKerr08} could also be considered a special case of such a greedy approach).  Both the greedy methods, and the one developed here, aim to optimize a similar objective function.  While greedy methods reduce the computational burden by restricting the search space of spike trains, here analytic approximations are made.  The advantage of the greedy approaches relative to this one is that they result in a spike train (ie, a binary sequence).  However, because of the numerical approximations and restrictions, one can never be sure whether the algorithm finds the \emph{most} likely possible spike train.  On the other hand, the approach developed herein is guaranteed to quickly find the most likely spike ``train'', but now the inferred spike train allows for partial spikes.  
One interesting future direction might be to explore whether greedy methods could be improved by initializing with a thresholded version of the \foopsi filter output.

Further, the \foopsi filter is based on a biophysical model capturing key features of the data, and may therefore be straightforwardly generalized in several ways to improve accuracy.  Unfortunately, some of these generalizations do not improve inference accuracy, perhaps because of the exponential approximation.  Instead, the \foopsi filter output can be used to initialize the more general SMC filter \cite{VogelsteinPaninski09}, to further improve inference quality (cf. Figure \ref{fig:smc_init}).  Another model generalization allows incorporation of spatial filtering of the raw movie into this approach (cf. Figure \ref{fig:spatial}).  

A number of extensions follow from this work.  First, pairing this filter with a crude but automatic segmentation tool to obtain ROIs would create a completely automatic algorithm that converts raw movies of populations of neurons into populations of spike trains.  Second, combining this algorithm with recently developed connectivity inference algorithms on this kind of data \cite{MishchenkoPaninski09}, could yield very efficient connectivity inference.

\paragraph{Acknowledgments}

The authors would like to express appreciation for helpful discussions with Vincent Bonin.  Support for JTV was provided by NIDCD DC00109. LP is supported by an NSF CAREER award, by an Alfred P.\ Sloan Research Fellowship, and the McKnight Scholar Award. RY's laboratory is supported by NIH EY11787 and the Kavli Institute for Brain Studies. LP and RY share a CRCNS award, NSF IIS-0904353.

\appendix

\section{Pseudocode} \label{sec:pseudo}

\begin{algorithm}[h!]
\caption{Pseudocode for inferring the approximately most likely spike train, given fluorescence data. Note that $\xi_i \ll 1$ for $i \in \{1,2\}$; the algorithm is robust to small variations in each. The equations listed below refer to the most general equations in the text (simpler equations could be substituted when appropriate).  Curly brackets, $\{ \cdot \}$, indicate comments.}
\label{eqn:pseudocode}
\begin{algorithmic}[1]
\STATE initialize parameters, $\bth$ (section \ref{sec:init})
	
\WHILE{convergence criteria not met}
  \FOR[interior point method to find $\hbC$]{$z=1,0.1,0.01,\ldots, \xi_1$}
    \STATE Initialize $n_t=\xi_2$ for all $t=1,\ldots, T$, $C_1=0$ and $C_t = \gam C_{t-1} + n_t$ for all $t=2,\ldots, T$.
	\STATE let $\bC_{z}$ be the initialized calcium, and $\mP_{z}$, be the posterior given this initialization
	\WHILE[Newton-Raphson with backtracking line searches]{$\mP_{z'}< \mP_{z}$}
		\STATE compute $\bg$ using Eq.~\eqref{eq:g2}
		\STATE compute $\bH$ using Eq.~\eqref{eq:H2}
		\STATE compute $\bd$ using $\bH \backslash \bg$ \COMMENT{(block-) tridiagonal Gaussian elimination}
		\STATE let $\bC_{z'}=\bC_z + s \bd$, where $s$ is between $0$ and $1$, and $\mP_{z'}>\mP_z$ \COMMENT{backtracking line search}
	\ENDWHILE
  \ENDFOR
\STATE check convergence criteria
\STATE update $\valpha$ using Eq.~\eqref{eq:valpha}  \COMMENT{only if spatial filtering} 
\STATE update $\vbeta$ using Eq.~\eqref{eq:vbeta}
\STATE let $\sig$ be the root-mean square of the residual
\STATE let $\lam=\frac{1}{T}\sum_t \hn_t$
\ENDWHILE
\end{algorithmic}
\end{algorithm}

\section{Wiener Filter} \label{sec:wiener}

The Poisson distribution in Eq.~\eqref{eq:n} can be replaced with a Gaussian instead of a Poisson distribution, ie,  $n_t \overset{iid}{\sim} \mN(\lam \Del, \lam \Del)$, which, when plugged into Eq.~\eqref{eq:nhat2} yields:
\begin{align} \label{eq:obj5}
\hbn &= \argmax_{n_t}  \sum_{t=1}^T \bigg( \frac{1}{2 \sig^2}(F_t - \alpha(C_t + \beta))^2  + 
 \frac{1}{2 \lam \Del}(n_t - \lam \Del)^2\bigg).
\end{align}
Note that since fluorescence integrates over $\Delta$, it makes sense that the mean scales with $\Delta$.  Further, since the Gaussian here is approximating a Poisson with high rate \cite{SjulsonMiesenbock07}, the variance should scale with the mean.  Using the same tridiagonal trick as above, Eq.~\eqref{eq:obj3} can be solved using Newton-Raphson once (because this expression is quadratic in $\bn$).  Writing the above in matrix notation and substituting $C_t - \gam C_{t-1}$ for $n_t$, yields:
\begin{align}   \label{eq:w2}
\hbC&= \argmax_{\bC} -\frac{1}{2\sig^2} \norm{\bF - \bC}^2 - \frac{1}{2\lam\Del} \norm{\bM \bC - \lam\Del\ve{1}}^2,
\end{align}
\noindent which is quadratic in $\bC$.  The gradient and Hessian are given by:
\begin{align}
\bg &= -\frac{1}{\sig^2} (\bC - \bF) - \frac{1}{\lam\Del} ( (\bM \hbC)\T \bM + \lam\Del \bM\T \ve{1}), \\
\bH &= \frac{1}{\sig^2} \ve{I} + \frac{1}{\lam\Del} \bM\T \bM.
\end{align}
Note that this solution is the optimal linear solution, under the assumption that spikes follow a Gaussian distribution, and is often referred to as the Wiener filter, regression with a smoothing prior, or ridge regression \cite{CONV04}.  Estimating the parameters for this model follows similarly as described in section \ref{sec:learn}.

\clearpage
\bibliography{biblist}
\addcontentsline{toc}{section}{References}
\bibliographystyle{ieeetr}


\end{document}